\documentclass[12pt]{article}
\usepackage{latexsym}
\usepackage{amssymb}
\usepackage{graphicx,epsfig}
\usepackage{pdfpages}
\usepackage{relsize}
\usepackage{graphicx}
\usepackage{amsfonts}
\usepackage{amsmath}
\usepackage{lipsum}
\usepackage{booktabs}
\RequirePackage[round,sort&compress]{natbib}
\RequirePackage[colorlinks,citecolor=blue,urlcolor=blue,linkcolor=blue]{hyperref}
\usepackage{authblk}
\makeatletter
\newcommand{\singlespacing}{\let\CS=\@currsize\renewcommand{\baselinestretch}{1}\tiny\CS}
\newcommand{\oneandahalfspacing}{\let\CS=\@currsize\renewcommand{\baselinestretch}{1.25}\tiny\CS}
\newcommand{\doublespacing}{\let\CS=\@currsize\renewcommand{\baselinestretch}{1.35}\tiny\CS}

\def\@citex[#1]#2{\if@filesw\immediate\write\@auxout{\string\citation{#2}}\fi
  \def\@citea{}\@cite{\@for\@citeb:=#2\do
    {\@citea\def\@citea{,\linebreak[0]\hskip0pt plus .2em}%
      \@ifundefined{b@\@citeb}%
    {{\bf ?}\@warning{Citation `\@citeb' on page \thepage\space undefined}}%
      \hbox{\csname b@\@citeb\endcsname}}}{#1}}

\date{}
\RequirePackage{graphicx}
\usepackage{float}
\RequirePackage[round,sort&compress]{natbib}
\RequirePackage[colorlinks,citecolor=blue,urlcolor=blue,linkcolor=blue]{hyperref}
\begin{document}

\title{\bf  Isotropic uncharged model with compactness and stable configurations}
\author[1]{\bf Amit Kumar Prasad \thanks{amitkarun5@gmail.com} }
\author[1]{\bf Jitendra Kumar\thanks{ jitendark@gmail.com}}   
\author[2]{\bf Ashok Kumar\thanks{ashrsdma@gmail.com} }
\affil[1]{\small Department of Mathematics, Central University of Jharkhand,Ranchi-835205 India.}
\affil[2]{\small Department of Mathematics, Hemwati Nandan Bhuguna University, Srinagar (Uttarakhand), India.}
\date{}
\maketitle \setlength{\parskip}{.11in}
\setlength{\baselineskip}{15pt}




\section*{Abstract}
In present work, we have studied a new stellar distribution model with spherically symmetric matter and an uncharged isotropic distribution in general relativity. In this model we have considered a particular metric potential. The model is capable to represents some known compact stars like Her X-1,4U 1538-52 and SAX J1808.4-3658.The model satisfy the energy condition and hydrostatic equilibrium equation, i.e., the modified Tolman-Oppenheimer-Volkoff (TOV) equation for uncharged matter. In addition to this,we also present the velocity of sound, surface redshift and pressure density ratio. The physical quantities such as pressure,density, redshift etc.,are compared with graphical representations that are important from theoretical and astrophysical scale.\\ \vspace{1cm}
\textbf{Keywords:}Isotropic Fluids; Compact star; General Relativity.
\section{Introduction}
An analysis of the solution of Einstein field equation shows that the exact solution plays an important role in development of many areas of gravitational field such as black hole solution,solar system test,gravitational collapse and so on. Generally in astrophysics compact stars formed due to gradual gravitational collapse are considered to fall the three categories, white dwarfs, neutron stars and black holes. This classification is based on the internal structure and composition of stars, where the formers contain matter in one of the densest forms found in the universe. According the strange matter hypothesis strange quark matter could be more stable than nuclear matter and thus neutrons star should largely be composed of pure quark matter. Possible observational signatures associated with theoretically proposed states of matter inside the compact stars therefore have remained an active research area in astrophysics , different types of mathematical modeling of such compact objects being considered.The singularity free interior solutions of compact object have an important consequences in relativistic astrophysics.The study of high density object like neutron stars,quark stars and white dwarfs,form their microscopic composition and properties of dense matter is one of the most fundamental problem in modern astrophysics.\par In general, it is important to measure the mass and radius\cite{Burikham,Boehmer1} of compact stars which depends on the equation of state \cite{Ray1,Negreiros,Varela,Maharaj1}.The motivation to undertake such a task, because the interior structure of such compact stars can vary with mass.On other hand, Buchdahl\cite{Buchdahl} proposed a method on the mass radius ratio of relativistic fluid spheres which is an important contribution in order to study the stability of the fluid spheres.Thus the motivation of this study is to prediction of mass and radius of compact stars.The mass and radius of compact stars such as Her X-1,4U 1538-52 and SAX J1808.4-3658 has been analyzed by Gangopadhyay et al\cite{gangopadhya}.\par The exact solution of Einstein-Maxwell field equations for static isotropic astrophysical object is of continuous interest to mathematician as well as physicists\cite{amit,amit1}. A large number of solutions have studied in\cite{Pratibha,Bhar,Bhar1,Takisa,Maurya1,Maurya2,Maurya4,Lemos,Kouretsis,Rahaman,Hansraj,YK,Esculpi,2013,2014} for exact solution of Einstein-Maxwell field equation.Some pioneer work in relativity is given by Ivanov\cite{Ivanov},Ray et al.\cite{Ray},Stettner\cite{stettner},Krori and Barua\cite{krori},Ray and Das\cite{das},Pant and Negi\cite{pant}, Florides\cite{Florides}, Dionysiou\cite{Dionysiou},Pant et al.\cite{pant1} etc.gave well behaved solution for charged fluid sphere.\par The equation of state(EOS) is an important features to describe a self gravitating fluid when it comes to solving the field equations. Ivanov\cite{Ivanov} has showed that the analytical solutions in the static, spherically symmetric uncharged case of perfect fluid with linear EOS is an extremely difficult problem. Sharma and Maharaj\cite{maharaj} have demonstrated this complexity in the case of a static, spherically symmetric uncharged anisotropic fluid. In the resent model we choose Buchdahl metric and solve the system of field equations and obtained a linear EOS. 
\par The algorithm for an anisotropic uncharged fluid has been done by Lake and Herrera\cite{Lake,Klake,Herrera}.In this work Lake\cite{Lake,Klake} has considered an algorithm and choose a single monotonic function which is generates a static spherically symmetric perfect fluids solutions of Einstein's equation.Against the above studies we have considered an uncharged isotropic fluid distribution in the context of the formation of the compact stars and find a new solution in section 2, section 3 consists of physical conditions for well behaved solutions.In section 4 the matching condition of interior metric to an exterior Reissner-Nordstrom line element and determine the constant coefficient. Stability analysis of compact objects and for better illustration of our result, the relevant physical quantities are presented by table and figure in section 5.Finally in section 6 we have drawn contains about present model.
~~~~~~~~~~~~~~~~~~~~~~~~~~~~~~~~~~~~~~~~~~~~~~~~~~~~
\section{Field equations for Uncharged Fluid Sphere in Schwarzschild Coordinates}
Let us consider the spherically symmetric metric in Schwarzschild Coordinates  \begin{eqnarray}
 ds^{2}=-e^{\lambda(r)}dr^{2} -r^{2}(d\theta^{2}+\sin^{2}\theta d\phi^{2})+e^{\nu(r)}dt^{2}\label{1}
\end{eqnarray} 
where $\lambda(r)$ and $\nu(r)$ are the functions of $ r  $ only.The Einstein equation for a perfect fluid distribution is given by  \begin{eqnarray}R^{i}_{j}-\dfrac{1}{2}R\delta^{i}_{j}=-\kappa \big[(c^{2}\rho+p)\nu^{i}\nu_{j}-p\delta^{i}_{j}\big]\label{2}
\end{eqnarray}
where $ \kappa=\dfrac{8\pi G}{c^{4}}$,with $G$ and $c$ are the gravitational constant and speed of light
in vacuum respectively. Here $\rho$ and  $ p $ denote matter density and fluid pressure respectively. The $\nu^{i}$ is the time-like 4-velocity vector such that \begin{eqnarray*} \nu^{i}\nu_{i}=1 \label{3}
\end{eqnarray*}\par
In view of the metric (\ref{1}) the Einstein field equations are given by \begin{eqnarray}
\dfrac{\nu'}{r}e^{-\lambda}-\dfrac{(1-e^{-\lambda})}{r^{2}}=\kappa p\label{4} \\
\bigg(\dfrac{\nu''}{2}-\dfrac{\lambda'\nu'}{4}+\dfrac{\nu^{'2}}{4}+\dfrac{\nu'-\lambda'}{2r}\bigg)e^{-\lambda}=\kappa p\label{5}\\
\dfrac{\lambda'}{r}e^{-\lambda}+\dfrac{(1-e^{-\lambda})}{r^{2}}=\kappa c^{2}\rho\label{6}
\end{eqnarray}
where prime $(')$ denotes the differentiation with respect to $r$. Now we consider a well known form of metric potential,which was proposed by Buchdahl\cite{Buchdahl} of the form as \begin{eqnarray} e^{\lambda}=\dfrac{K(1+Cr^{2})}{K+Cr^{2}},~~~~K>1\label{7} \end{eqnarray} where $ K $ and $ C $ are arbitrary constant.The metric function(\ref{7}) is regular and non-singular at the center of the star which satisfies the primary physical requirements for a realistic star.
 Now using (\ref{7}),the equations (\ref{4})-(\ref{6}) reduce to the following form
	\begin{eqnarray}\dfrac{(K+Cr^{2})}{K(1+Cr^{2})}\bigg[\dfrac{-2y'}{ry}+\dfrac{C(K-1)}{K+Cr^{2}}\bigg]=-\kappa p \label{8}\\	\dfrac{C(K-1)(3+Cr^{2})}{K(1+Cr^{2})^{2}}=\kappa c^{2}\rho\label{9} \\ 
	\dfrac{(K+Cr^{2})}{K(1+Cr^{2})}\bigg[\dfrac{y''}{y}-\dfrac{y'}{ry}+\dfrac{C(K-1)r(Cr-y'/y)}{(K+Cr^{2})(1+Cr^{2})} \bigg]=0\label{10}
	\end{eqnarray}
where $ e^{\nu} =y^2$. Now to solve the equation (\ref{10}) we introduce the new variables define by \begin{eqnarray}
  X=\sqrt{\dfrac{K+Cr^{2}}{K-1}},~~~~K>1 \label{11}
 \end{eqnarray}and \begin{eqnarray}
  y(X)=(X^{2}-1)^{1/4}\Psi(X)\label{12}
  \end{eqnarray}Using the equations (\ref{11}) and (\ref{12}), the equation(\ref{10}) reduce to the following form of second order differential equation \begin{eqnarray}
    \dfrac{d^{2}\Psi}{dX^{2}}+I\Psi=0\label{13}
  \end{eqnarray} where \begin{eqnarray} 
    I=\dfrac{2(1-2K)(1-X^2)-5X^2}{4(1-X^{2})^2}\label{14}
 \end{eqnarray} In order to solve equation (\ref{13}) more easily if we set $ K=\frac{7}{4} $, then the equation (\ref{13}) takes the form \begin{eqnarray}
   \Psi''-\dfrac{5}{4(1-X^{2})^{2}}\Psi=0\label{15}
 \end{eqnarray}
 So the solution of (\ref{15}) is given as
 \begin{eqnarray}
 \Psi(X)=(X+1)\left[A_{1}\bigg\rvert\dfrac{X+1}{X-1}\bigg\rvert^{(0.25)}  +A_{2}\bigg\rvert\dfrac{X+1}{X-1}\bigg\rvert^{(-1.25)} \right]\label{16}
 \end{eqnarray} where $ A_1 $ and $ A_2 $ are arbitrary constants of integration. Now put the value of $ \Psi(X) $ from equation (\ref{16}) and $ X=\sqrt{\dfrac{K+Cr^{2}}{K-1}} $ into the equation (\ref{12}),we get \begin{eqnarray}
 y(r)=\left[\dfrac{4(1+Cr^{2})}{3}\right]^{1/4}\Big(g(r)+1\Big)\Bigg[A_{1}\Big\rvert F(r)\Big\rvert^{(0.25)} +A_{2}\Big\rvert F(r)\Big\rvert^{(-1.25)} \Bigg]\label{17} \end{eqnarray}
 where $ g(r)=\sqrt{\frac{7+4Cr^2}{3}},~~~F(r)=\dfrac{5+2Cr^{2}+3g(r)}{2(1+Cr^2)} $ . Therefore the expressions of density and pressure are given by \begin{eqnarray}
 \kappa c^{2}\rho=\dfrac{3C(3+Cr^{2})}{7(1+Cr^{2})^{2}}\label{18}\end{eqnarray} 
 \begin{eqnarray}
 \kappa p=\dfrac{2(7+4Cr^2)}{7ry(r)(1+Cr^2)}\Bigg[ N_{1}(r)\Bigg(A_{1}\Big\rvert F(r)\Big\rvert^{(0.25)} +A_{2}\Big\rvert F(r)\Big\rvert^{(-1.25)}\Bigg)+N_{2}(r)N_{3}(r)N_{4}(r)\Bigg]-\dfrac{3C}{7(1+Cr^2)} \label{19}
\end{eqnarray}
where \\$
N_{1}(r)=\dfrac{2Cr}{3}
\Bigg[\big(g(r)+1\big)\Bigg(\frac{4(1+Cr^2)}{3}\Bigg)^{-3/4}+2\Bigg(\frac{4(1+Cr^2)}{3}\Bigg)^{1/4}g(r)\Bigg],~~~~~~~N_{4}(r)=\dfrac{-8Cr}{3g(r)\big(g(r)-1\big)^2} \\
N_{2}(r)=\Bigg(\frac{4(1+Cr^2)}{3}\Bigg)^{1/4}\big(g(r)+1\big),~~~~~~~~~~~N_{3}(r)=\Bigg[A_{1}(0.25)\Big\rvert F(r)\Big\rvert^{(-0.75)} +A_{2}(-1.25)\Big\rvert F(r)\Big\rvert^{(-2.25)} \Bigg]\\
$
\begin{figure}[h]
\begin{center}
\includegraphics[width=8cm]{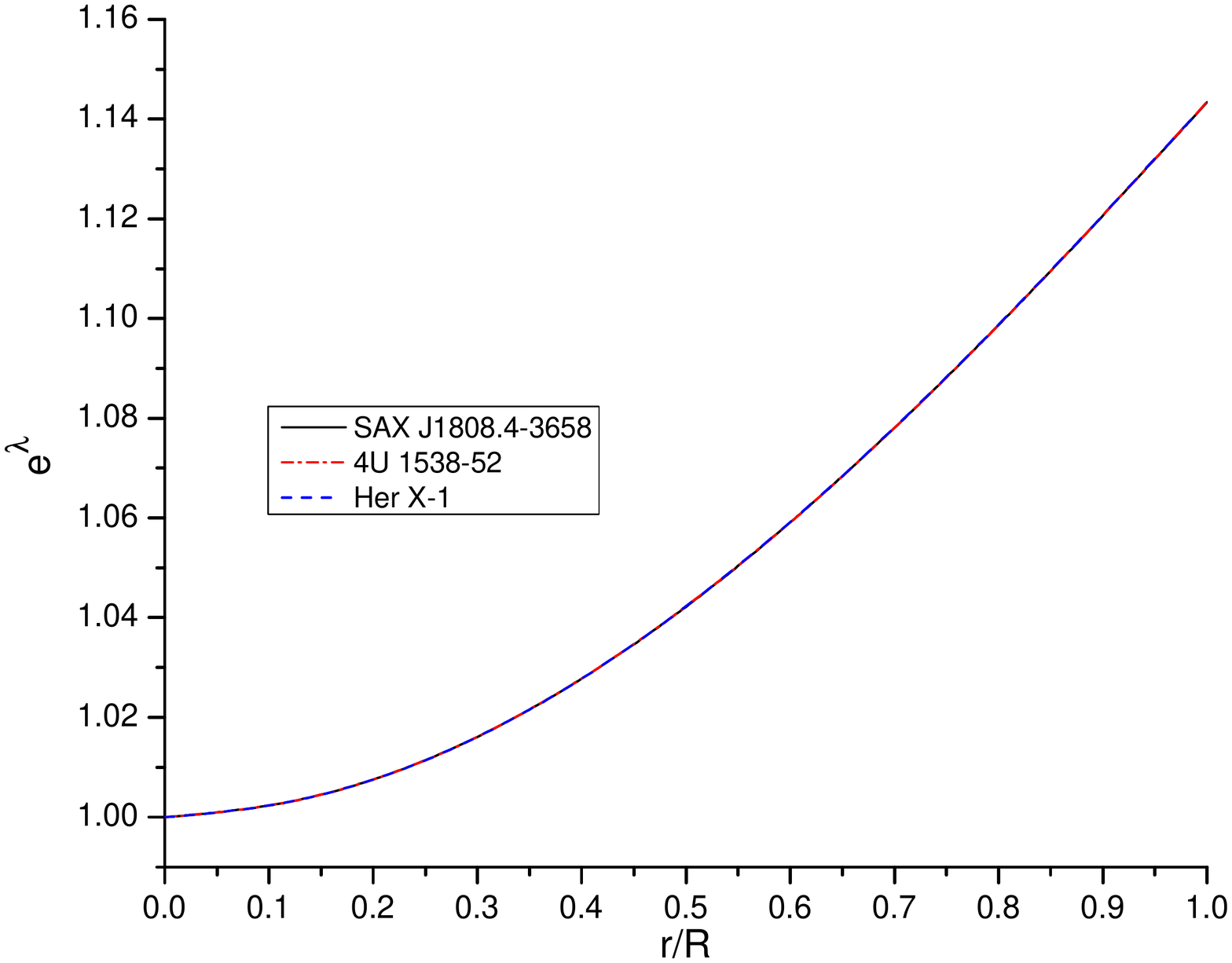}\includegraphics[width=8cm]{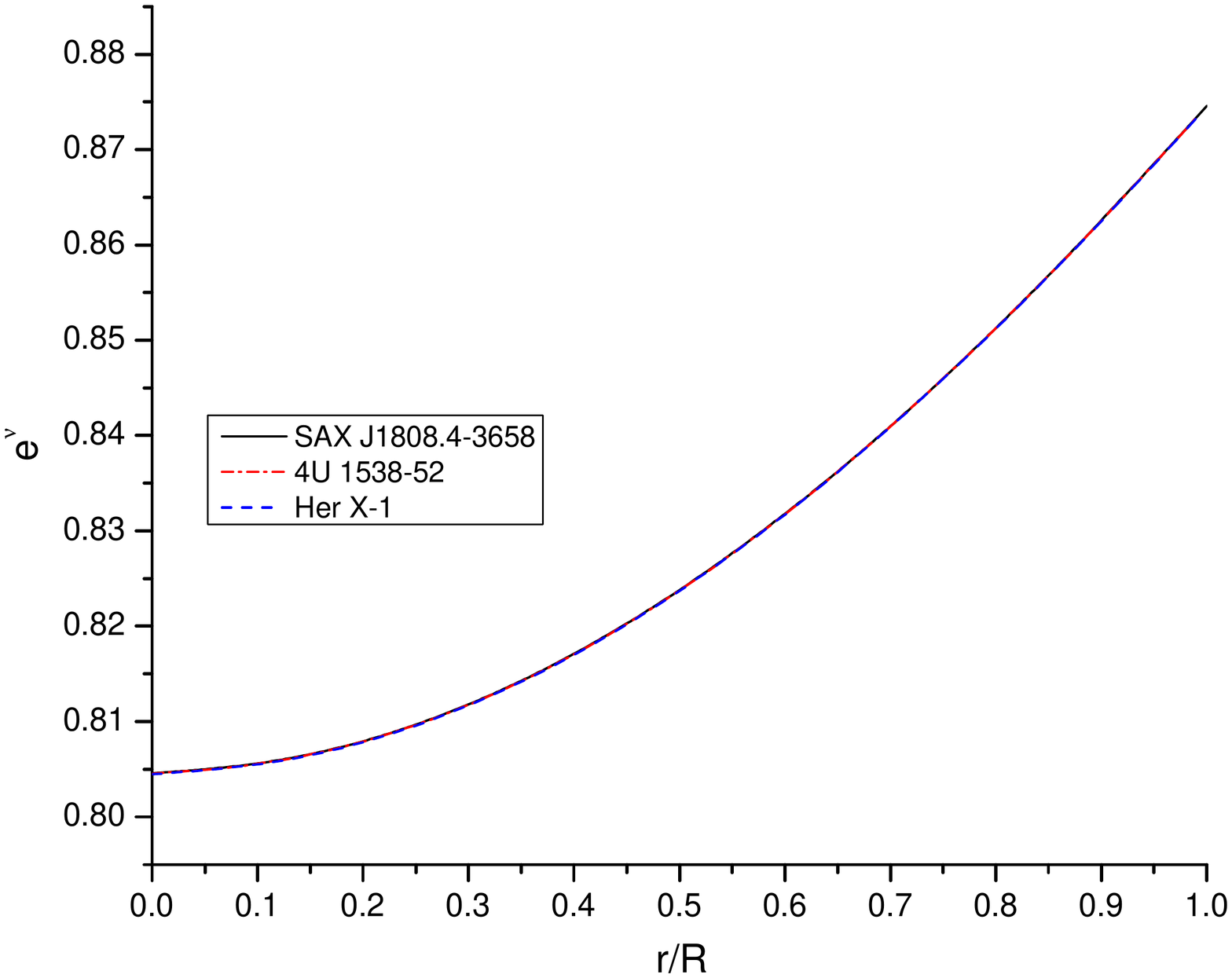}
\caption{Variation of metric potential $ e^{\lambda} $(left) and $ e^{\nu} $(right) with respect to fractional radius $ r/R $ for compact star SAX J1808.4-3658,4U 1538-52 and Her X-1. For plotting this figure the numerical values of physical parameters and constants are as follows: (i)$ K = 1.75, CR^2 =0.4137, M$ =$0.9M_{\odot}$ and $R =14.35\,km $ for SAX J1808.4-3658,(ii)$ K =1.75,CR^2=0.4138,M =$ $0.87M_{\odot}$ and $R =13.87\, km$ for 4U 1538-52 ,(iii)$ K =1.75,CR^2=0.414,M =$ $0.85M_{\odot}$ and $R =13.548\, km$ for Her X-1 }\label{f1}
\end{center}
\end{figure}
\section{Physical Features for Well Behaved Solution}
\begin{figure}[h!]
\begin{center}
\includegraphics[width=6cm]{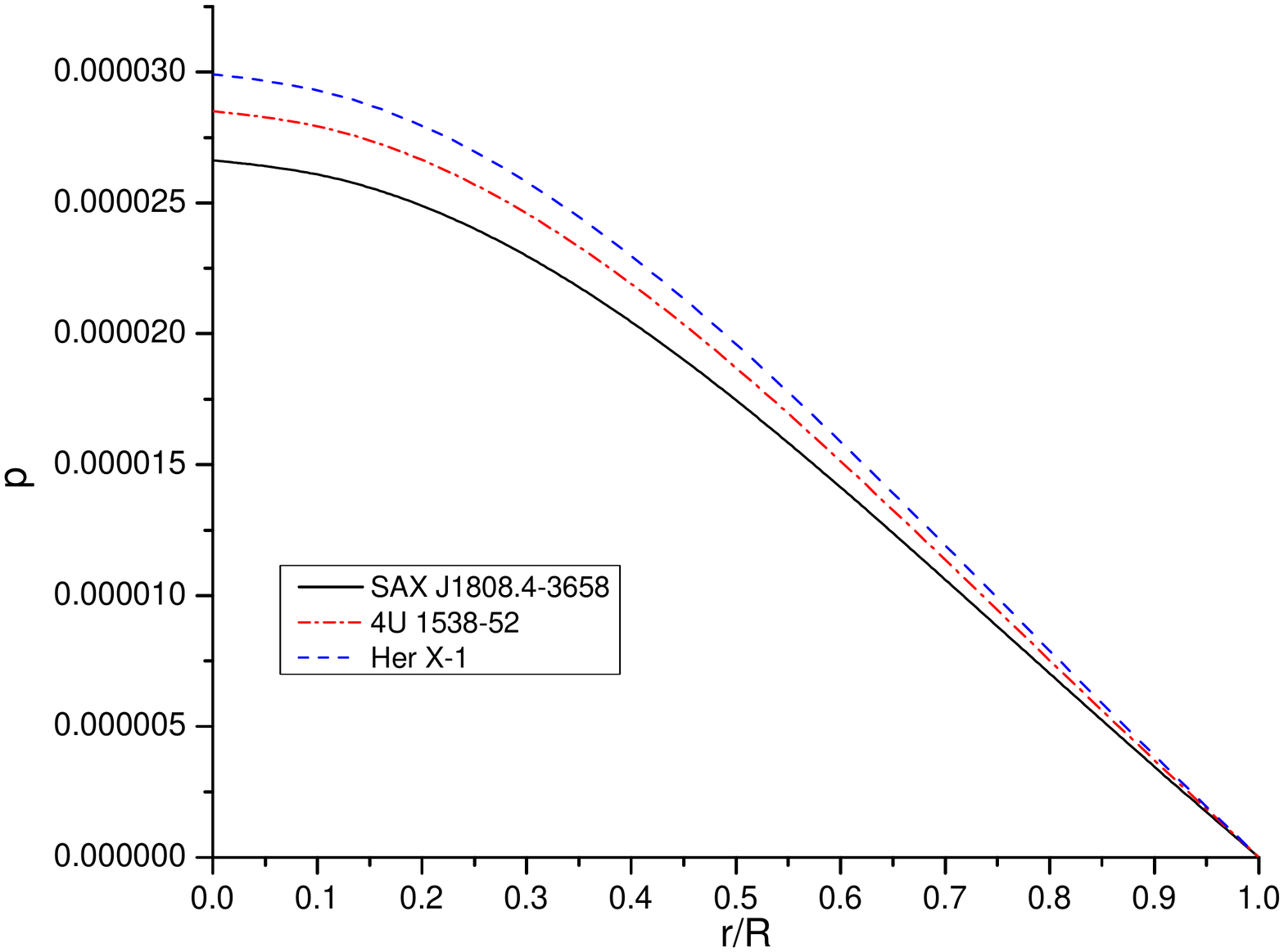}\includegraphics[width=6cm]{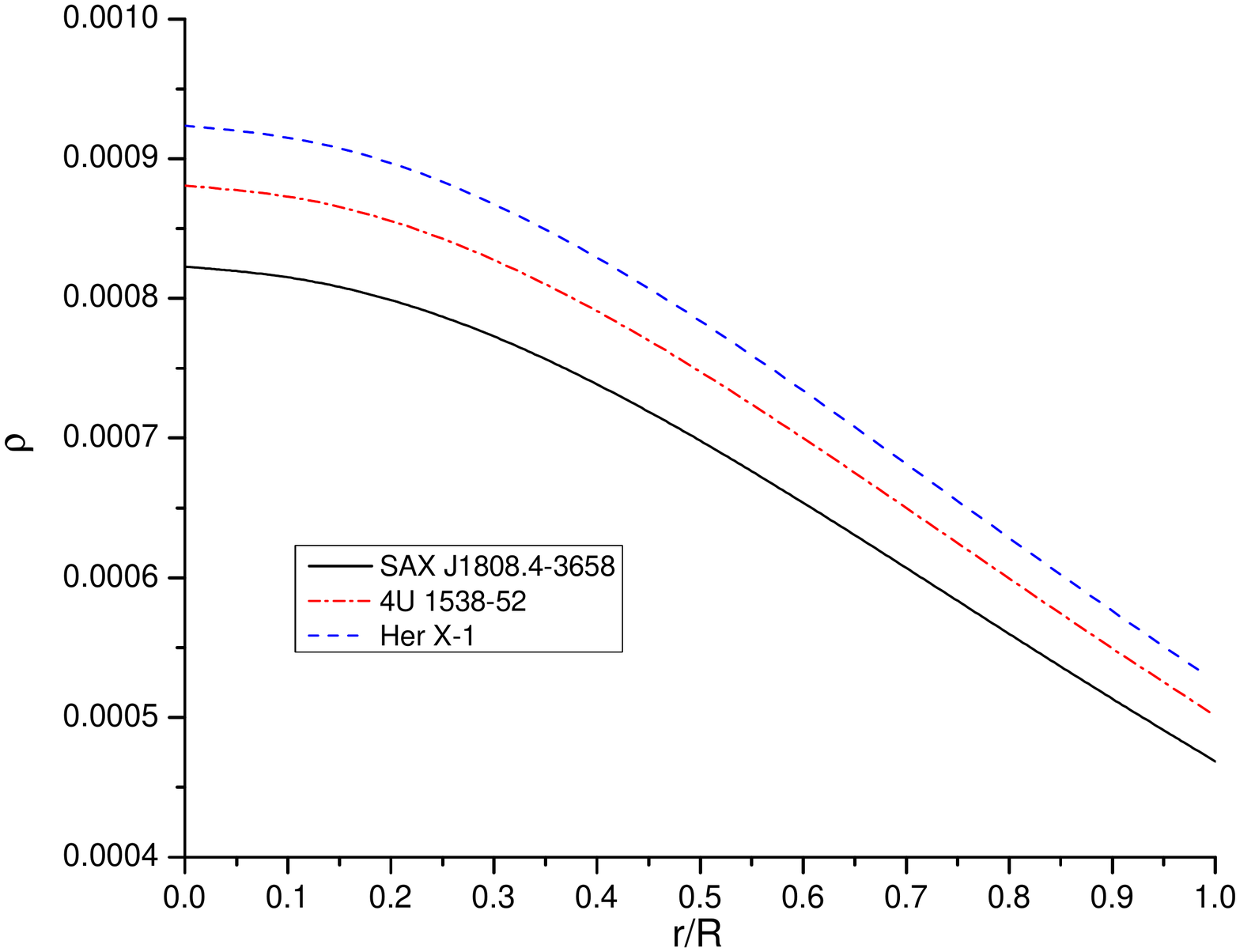}\includegraphics[width=6cm]{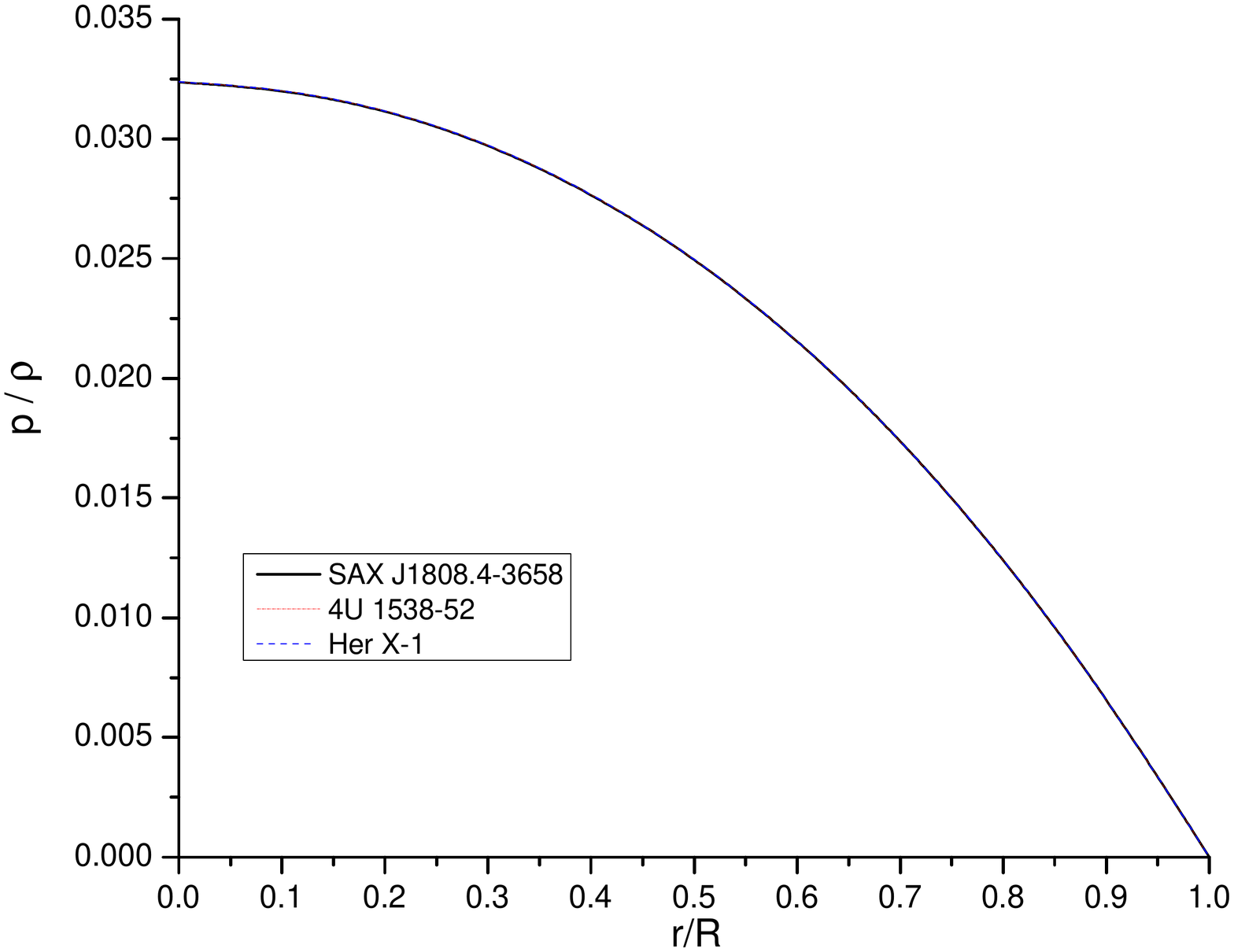}
\caption{Behaviour of pressure(p in $ km^{-2} $ left), density($ \rho $ in $ km^{-2} $ middle) and $ p/\rho $(right) vs. fractional radius r/R for SAX J1808.4-3658, 4U 1538-52 and Her X-1.
For plotting this figure the numerical values of physical parameters and constants with $ G=c=1 $ are as follows: (i)$ K = 1.75, CR^2 =0.4137, M$ =$0.9M_{\odot}$ and $R =14.35\, km$ for SAX J1808.4-3658,(ii)$ K =1.75,CR^2=0.4138,M =$ $0.87M_{\odot}$ and $R =13.87\, km$ for for 4U 1538-52 ,(iii)$ K =1.75,CR^2=0.414,M =$ $0.85M_{\odot}$ and $R =13.548\, km$ Her X-1}\label{f2}
\end{center}
\end{figure}

\begin{enumerate}
\item From equation (\ref{7}),we observe $ (e^{\lambda})_{(r=0)}=1 $ and $ (e^{\nu})_{(r=0)}>0 $. This show that metric potentials are singularity free and positive at center.It is monotonically increasing with increasing the radius of the compact star(see Fig.\ref{f1}).
\item Pressure $p$ should be zero at the boundary $ r=R. $
\item $ (dp/dr)_{r=0}=0 $ and $ (d^{2}p/dr^{2})_{r=0}<0, $ so that pressure gradient $ dp/dr $ is negative for $ 0<r \leq R $.
\item $ (d\rho/dr)_{r=0}=0 $ and $ (d^{2}\rho/dr^{2})_{r=0}<0, $ so that density gradient $ d\rho/dr $ is negative for $ 0<r \leq R $.\\
The above two conditions imply that pressure and density should be maximum at the center and monotonically decreasing towards the surface(see Fig.\ref{f2}).
\item The velocity of sound $ (dp/c^{2}d\rho)^{1/2} $ should be less than that of light throughout the charged fluid sphere $ (0\leq r\leq R) $. This is called casual condition. 
\item The ratio of pressure to the density $ (p/c^{2}\rho) $ should be monotonically decreasing with the increasing of $ r. $(see Fig.\ref{f2})
\item $ c^{2}\rho\geq p>0 $ or $ c^{2}\rho\geq 3p>0,0\leq r\leq R $,where former inequality denotes weak energy condition (WEC) and later inequality denotes strong energy condition (SEC).
\end{enumerate}
\section{Matching Conditions of Boundary }
The solution is smoothly connected to the pressure free boundary with the Schwarzschild exterior metric
 \begin{eqnarray}
 ds^{2}=-\bigg(1-\dfrac{2M}{r}\bigg)^{-1}dr^{2}-r^{2}(d\theta^{2}+\sin^{2}\theta d\phi^{2})+\bigg( 1-\dfrac{2M}{r}\bigg)dt^{2}\label{20} 
\end{eqnarray} 
  Beside the above the smooth joining  with the Schwarzschild metric which requires the continuity of $ e^{\lambda}$ and $e^{\nu} $  across the boundary  $ r=R $ and we get\\ \\ \begin{eqnarray} e^{-\lambda}=1-\dfrac{2M}{R}\label{24}\\ y^{2}=1-\dfrac{2M}{R}\label{25}\\p(R)=0.\label{27}
 \end{eqnarray}
 Using the equations (\ref{25}) and (\ref{27}),we get the expressions of arbitrary constant $ A_1 $ and $ A_2 $ as follow \begin{eqnarray}
 \dfrac{A_1}{A_2}=\dfrac{M_{1}(R)N_{1}(R)G(R)+M_{1}(R)N_{4}(R)N_{2}(R)G_{1}(R)-M_{2}(R)N_{2}(R)G(R)}{M_{2}(R)N_{2}(R)G_{2}(R)+M_{1}(R)N_{2}(R)N_{4}(R)G_{3}(R)-M_{1}(R)N_{1}(R)G_{2}(R)}\label{28}
 \end{eqnarray}
 \begin{eqnarray}
  A_{2}=\dfrac{\sqrt{\frac{7+4CR^2}{7(1+CR^2)}}}{\left[\dfrac{4(1+Cr^{2})}{3}\right]^{1/4}\Big(g(R)+1\Big)\Bigg[\frac{A_{1}}{A_{2}}\Big\rvert F(R)\Big\rvert^{(0.25)} +\Big\rvert F(R)\Big\rvert^{(-1.25)} \Bigg]}\label{29}
  \end{eqnarray}
The expression of mass $M$ is given as
\begin{eqnarray}
M=\frac{R}{2}\Bigg[ 1-\dfrac{7+4CR^2}{7(1+CR^2)}\Bigg]\label{30}
\end{eqnarray}
where\\$  g(R)=\sqrt{\frac{7+4CR^2}{3}},~~~F(R)=\dfrac{5+2Cr^{2}+3g(R)}{2(1+CR^2)},~~~~~~N_{4}(R)=\dfrac{-8}{3g(R)\big(g(R)-1\big)^2}\\~~~N_{1}(R)=(2/3)\Bigg[\big(g(R)+1\big)\Bigg(\frac{4(1+CR^2)}{3}\Bigg)^{-3/4}+2\Bigg(\frac{4(1+CR^2)}{3}\Bigg)^{1/4}g(R)\Bigg],~~~~~~~ M_{1}(R)=\dfrac{2C(7+4CR^2)}{7(1+CR^2)}\\
N_{2}(R)=\Bigg(\frac{4(1+CR^2)}{3}\Bigg)^{1/4}\big(g(R)+1\big),~~~~~~~G(R)=\big(F(R)\big)^{1/4} ~~~~G_{1}(R)=\frac{1}{4}\big(F(R)\big)^{-3/4},\\~~~~~~G_{2}(R)=\big(F(R)\big)^{-5/4}, ~~~~~G_{3}(R)=\frac{5}{4}\big(F(R)\big)^{-9/4},~~~~M_{2}(R)=\dfrac{3C}{7(1+CR^2)}     $
\section{Stability Analysis of Compact Objects}
In this section we have studied physical properties of interior of the fluid sphere and equilibrium conditions under different forces.
\subsection{Causality Condition} 
The speed of sound $ \dfrac{dp}{c^2d\rho} $ is less than the velocity of light.Here we fix $c=1$, and obtained speed of sound for uncharged fluid matter.Herrera\cite{lherrera} states that  for the stability the value of sound belongs to region of the interval $ 0<v^2=\frac{dp}{d\rho}<1 $ and should be monotonically decreasing away from the center. Now from the equations (\ref{18}) and (\ref{19}),we get expression of speed of sound
\begin{figure}[h]
\begin{center}
\includegraphics[width=10cm]{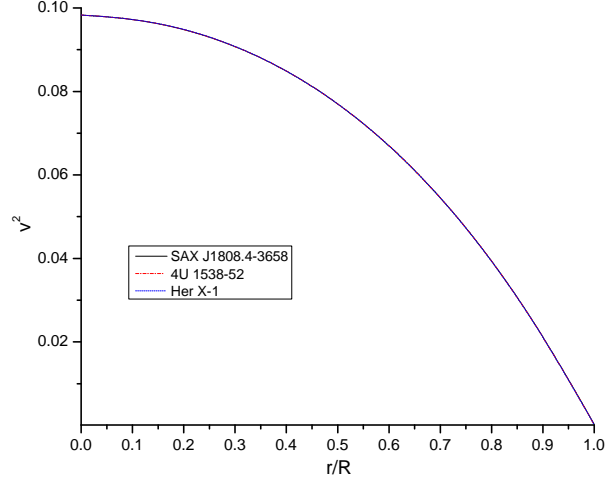}
\caption{Behaviour of velocity of sound vs. fractional radius r/R for SAX J1808.4-3658,4U 1538-52 and Her X-1.For plotting this figure we have employed data set values of physical parameters and constants which are the same as used in Fig.\ref{f2}}\label{f3}
\end{center}
\end{figure}
\begin{eqnarray}
\dfrac{dp}{d\rho}=\frac{L_{1}(r)\bigg(L_{2}(r)+L_{3}(r)\bigg)-L_{4}(r)L_{5}(r)-L_{6}(r)\bigg(L_{1}(r)\bigg)^2}{\bigg(L_{1}(r)\bigg)^2L_{7}(r)}\label{31}
\end{eqnarray}
where $\\
L_{1}(r)=\left[\dfrac{4(1+Cr^{2})}{3}\right]^{1/4}\Big(g(r)+1\Big)\Bigg[\Big\rvert F(r)\Big\rvert^{(0.25)} +\frac{A_{2}}{A_{1}}\Big\rvert F(r)\Big\rvert^{(-1.25)} \Bigg]\\
L_{2}(r)=\dfrac{-12C^2r}{7(1+Cr^2)^2}\Bigg[N_{1}(r) S(r)+N_{2}(r)S_{1}(r)N_{4}(r)\Bigg],~~~L_{4}(r)=\dfrac{2C(7+4Cr^2)}{7(1+Cr^2)}\Bigg[N_{1}(r)S(r)+N_{2}(r)S_{1}(r)N_{4}(r)  \Bigg]\\
L_{3}(r)=\dfrac{2C(7+4Cr^2)}{7(1+Cr^2)}\Bigg[ S_{2}(r)S(r)+2N_{1}(r)S_{1}(r)N_{4}(r)+N_{2}(r)S_{3}(r)S_{1}(r)+N_{2}(r)N_{4}(r)S_{4}(r)\Bigg]\\
L_{5}(r)=N_{1}(r)S(r)+N_{2}(r)S_{1}(r)N_{4}(r),~~~~ L_{6}(r)=\dfrac{-6C^2r}{7(1+Cr^2)^2},~~~~~~L_{7}(r)=\dfrac{-6C^2r(5+Cr^2)}{7(1+Cr^2)^3}\\S(r)=\Bigg[\Big\rvert F(r)\Big\rvert^{1/4} +\frac{A_{2}}{A_{1}}\Big\rvert F(r)\Big\rvert^{-5/4} \Bigg],~~~~S_{1}(r)=\Bigg[(0.25)\Big\rvert F(r)\Big\rvert^{-3/4} +\frac{A_{2}}{A_{1}}(-1.25)\Big\rvert F(r)\Big\rvert^{-9/4} \Bigg]\\ \\ \\
S_{2}(r)=\frac{8Cr}{9}\left[-\dfrac{6\big(g(r)+1\big)}{\Big(4\frac{4(1+Cr^2)}{3}\Big)^{7/4}}+\dfrac{2}{g(r)\Big(\frac{4(1+Cr^2)}{3}\Big)^{3/4}}-\dfrac{2{\Big(\frac{4(1+Cr^2)}{3}\Big)^{1/4}}}{\big(g(r)\big)^3}\right],~~~~~~S_{3}(r)=\dfrac{16Cr}{3g(r)\bigg(g(r)-1\bigg)^3}\\ S_{4}(r)=\Bigg[(-3/16)\Big\rvert F(r)\Big\rvert^{-7/4}-\frac{A_{2}}{A_{1}}(-45/16)\Big\rvert F(r)\Big\rvert^{-13/4} \Bigg]N_{4}(r)$\\
For better understanding we use the graphical representation to represent on Fig.(\ref{f3}).Thus,it is clear that in the Fig.(\ref{f3}), the speed of sound lies within the proposed interval and therefore this result maintains stability.
\subsection{Tolman-Oppenheimer-Volkoff (TOV) equations}
The general-relativistic hydrostatic equations were developed and used to models of compact stars by Tolman, Oppenhiemer and volkoff in \cite{35}. These equations are obtained from Einstein-Maxwell field equations when metric is static and isotropic.The latter hypothesis is predicted to be a well approximation for densest interior of static compact star, because the strong gravitational force is balanced by a huge pressure and rigid body forces have a negligible effect on the structure. In the connection of the microscopic theory for the relation between pressure and energy density, and the mass, this equation gives a equilibrium solution.The Tolman-Oppenheimer-Volkoff(TOV) equation \cite{35,36} in the presence of charge is given by
\begin{eqnarray}
 -\frac{M_G(\rho+p)}{r^2}e^{\frac{\lambda-\nu}{2}}-\frac{dp}{dr}+
 \sigma \frac{q}{r^2}e^{\frac{\lambda}{2}} =0,\label{32} 
 \end{eqnarray}
where $ \sigma  $ is charge density,$ q $ is charge and $M_G$ is the effective gravitational mass which is given by:
\begin{figure}[h]
\begin{center}
\includegraphics[width=6cm]{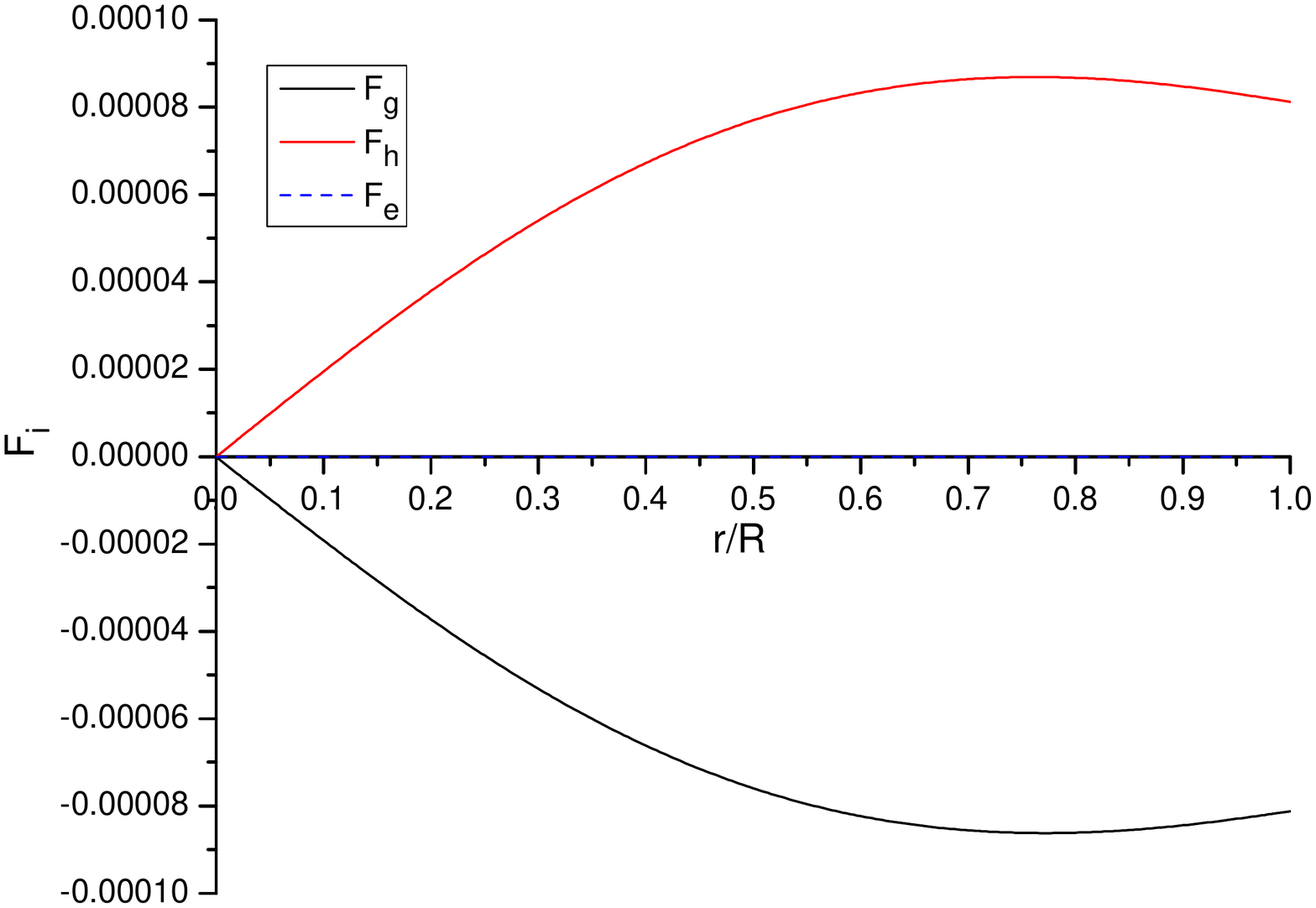}\includegraphics[width=6cm]{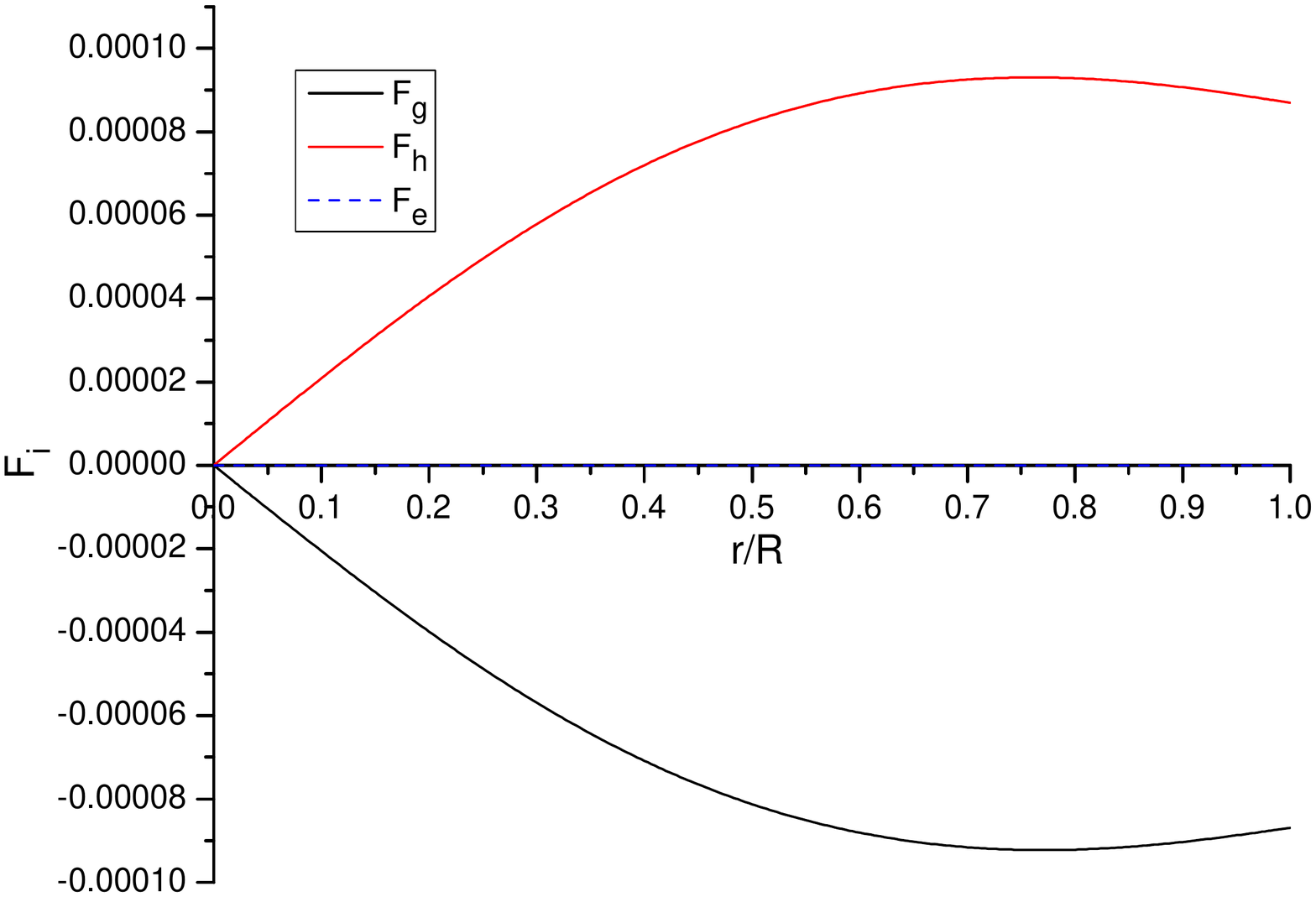}\includegraphics[width=6cm]{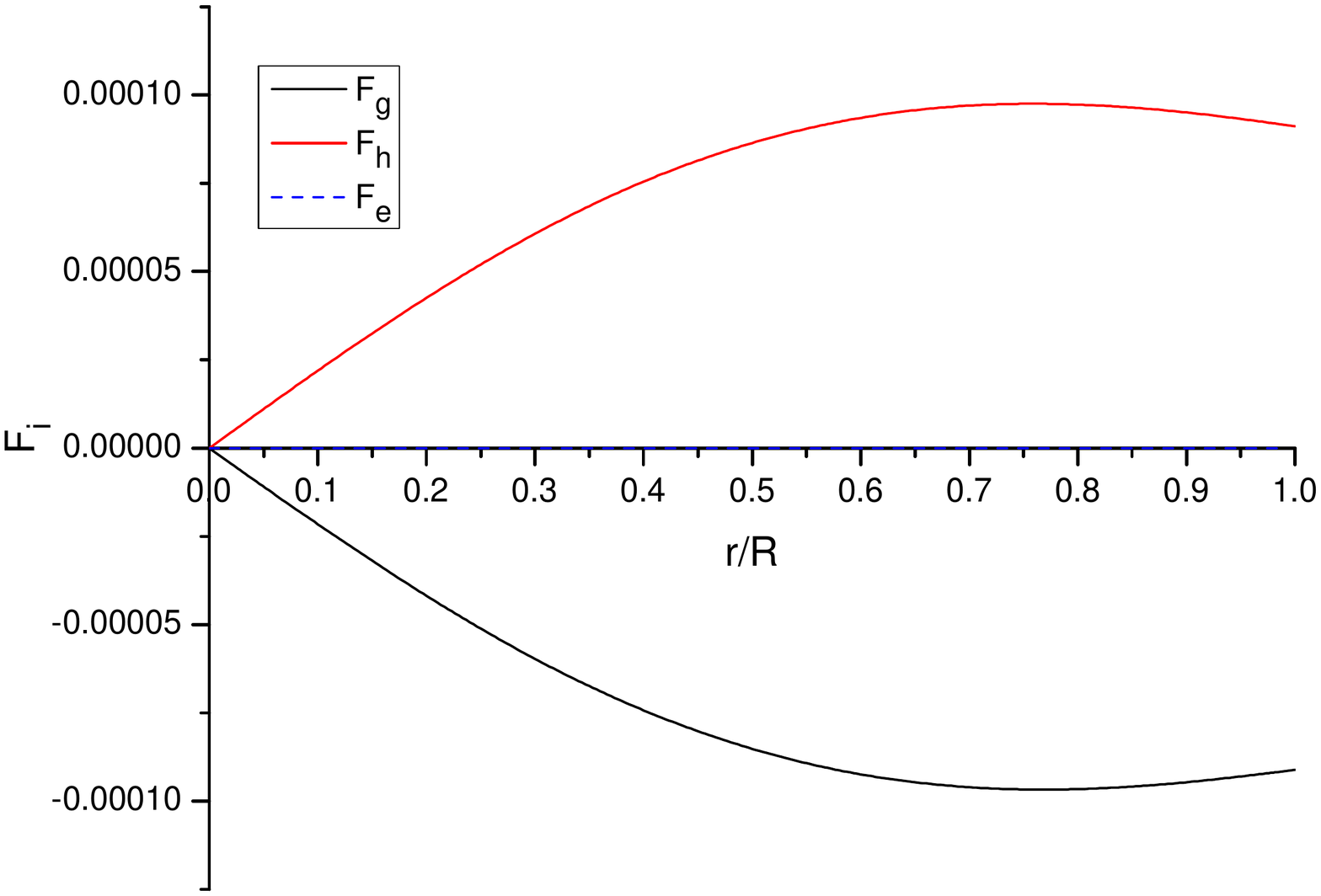}
\caption{Behaviour of different forces(in $ km^{-3} $ with $ G = c =1 $ ) vs. fractional radius $ r/R $. For plotting this figure the numerical values of physical parameters and constants are as follows:(i) $K =1.75,CR^2=0.4137,M =$ $0.9M_{\odot}$ and $R =14.35\, km$ for SAX J1808.4-3658(left), (ii)Her X-1(left) $ K = 1.74, CR^2 =0.4138, M$ =$0.87M_{\odot}$ and $R =13.87\, km$ for 4u 1538-52(middle),(iii) $ K = 1.74, CR^2 =0.414, M$ =$0.85M_{\odot}$ and $R =13.458\, km$ for Her X-1(left)}\label{f4}
\end{center}
\end{figure}
\begin{eqnarray}
M_G(r)=\frac{1}{2}r^2 \nu^{\prime}e^{(\nu - \lambda)/2}.\label{33}
\end{eqnarray}
Plugging the value of $M_G(r)$ in equation (\ref{32}), we get
\begin{eqnarray}
-\frac{\nu'}{2}(\rho+p)-\frac{dp}{dr}+\sigma \frac{q}{r^2}e^{\frac{\lambda}{2}} =0,  \label{34}
\end{eqnarray}
But in our model we have considered an uncharged isotropic fluid distribution i.e., the charge($ q $ ) is vanish so equation (\ref{34}) becomes
\begin{eqnarray}
-\frac{\nu'}{2}(\rho+p)-\frac{dp}{dr} =0,  \label{35}
\end{eqnarray}
The above equation can be expressed into three different components gravitational force $(F_g)$, hydrostatic force $(F_h)$ and electric force $(F_e)$, which are defined as: 
\begin{eqnarray}
F_g=-\frac{\nu'}{2}(\rho+p)=\dfrac{Z'}{8\pi Z}(\rho+p)\label{36}
\end{eqnarray}
\begin{eqnarray}
F_h=-\frac{dp}{8\pi dr}=-\frac{1}{8\pi}\left[
\frac{L_{1}(r)\bigg(L_{2}(r)+L_{3}(r)\bigg)-L_{4}(r)L_{5}(r)}{\bigg(L_{1}(r)\bigg)^2}-L_{6}(r) \right] \label{37}
\end{eqnarray}
\begin{eqnarray}
F_e=0\label{38}
\end{eqnarray}
where we use the same notation as above. Fig.(\ref{f4})  represents the behavior of the generalized
TOV equations. We observe from these figures that the system is counterbalanced by the components the gravitational force $(F_g)$ and hydrostatic force$(F_h)$ and the system attains a static equilibrium.
\subsection{Energy Condition}
 \begin{figure}[h]
\begin{center}
\includegraphics[width=6cm]{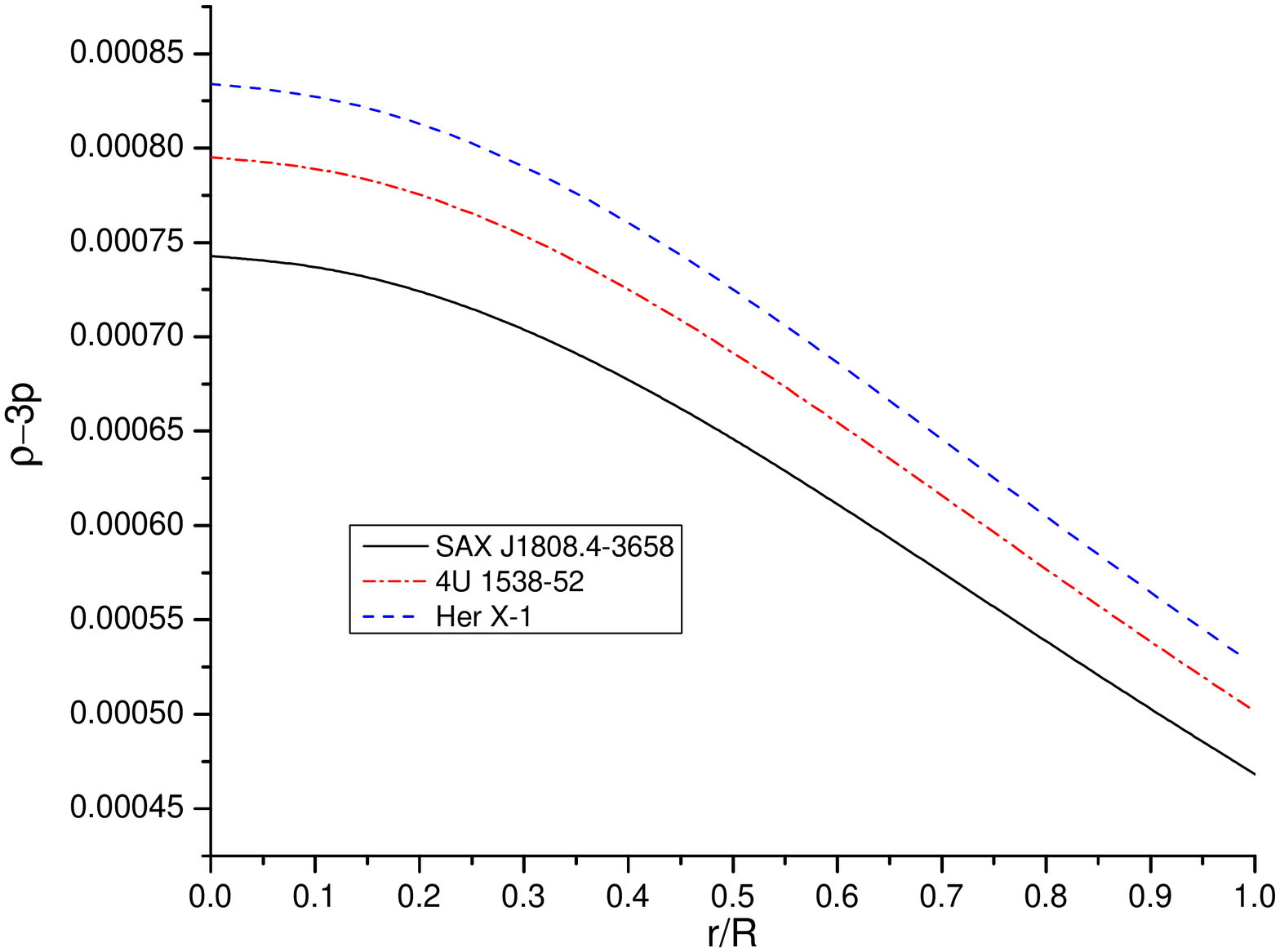}\includegraphics[width=6cm]{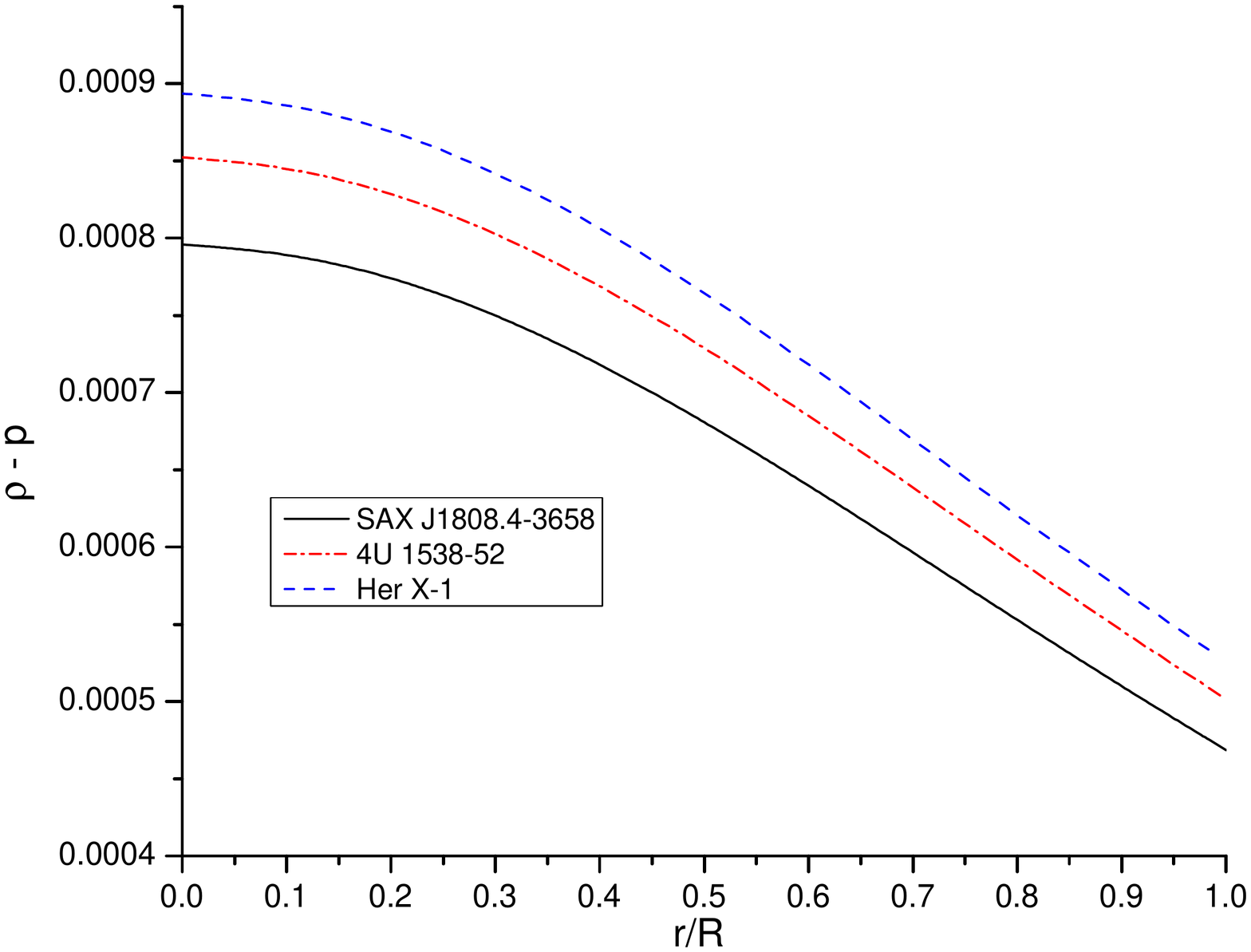}\includegraphics[width=6cm]{DEN.pdf}
\caption{Behaviour of different energy conditions(in $ km^{-2} $ with $ G = c =1 $) vs. fractional radius Curves are plotted for the SEC,WEC and NEC for the compact objects SAX J1808.4-3658, 4U 1538-52 and Her X-1. For plotting
this figure we have employed data set values of physical parameters and constants which are the same as used in Fig.\ref{f2} }\label{f5}
\end{center}
\end{figure}
The energy conditions depend on the matter density and pressure.That follow certain restrictions.Basic information about the energy condition in\cite{Pratibha}.Here we focus on the (i) Null energy condition, (ii) Weak energy conditions and(ii)Strong energy condition , which have the following inequalities
\begin{eqnarray}
\rho \geq 0\label{39}\\
\rho-p \geq 0\label{41}\\
\rho-3p \geq 0\label{42}
\end{eqnarray}
Using these inequalities we justify the nature of energy conditions for the specific stellar configuration as shown in Fig.(\ref{f5}), that satisfy our result.
\subsection{Adiabatic index}
\begin{figure}[h]
\begin{center}
\includegraphics[width=8cm]{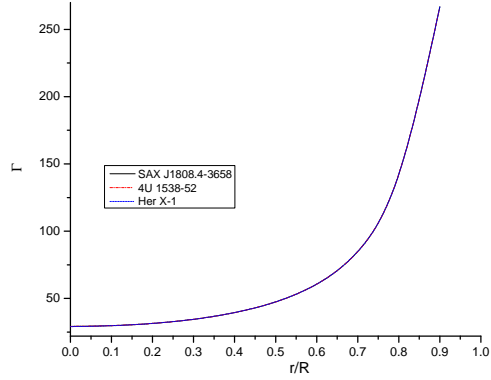}
\caption{Behaviour of adiabatic constant$(\Gamma)$ vs. fractional radius
r/R for SAX J1808.4-3658,4U 1538-52 and Her X-1. For plotting this figure we have employed data set values of physical parameters and constants which are the same as used in Fig.\ref{f2} }\label{f6}
\end{center}
\end{figure}

In order to have an equilibrium configuration the matter must be stable against the collapse of local regions. This requires Le Chateliers principle, also known as local or microscopic stability condition, that the pressure must be monotonically decreasing function of $r$ such that $\dfrac{dp}{d\rho}\geq 0.$ Heintzmann and Hillebrandt\cite{Heint} also proposed that compact star with the equation of state are stable for adiabatic index $\Gamma=\big(\frac{p+\rho}{p}\big)\frac{dp}{d\rho} > 4/3.$ Fig.\ref{f6} show that $\Gamma > 4/3$, so model developed in this paper is stable.
\begin{table}[h]
\caption{Numerical value of parameters $ C, M _{\odot}, R $ and Buchdahl limit for different compact stars.}
\label{T1}
\begin{tabular}{ccccc}
\hline\rule[-1ex]{0pt}{3.5ex}
Compact star candidates& M($M _{\odot}$) & Predicted radius $ R(km) $ & $ C(km^{-2}) $ & $ 2M/R\leq 8/9 $ \\
\hline\rule[-1ex]{0pt}{3.5ex} 
SAX J1808.4-3658&  0.9 & 14.35 &$ 2.009\times 10^{-13}$ & 0.18501\\
4U 1538-52 & 0.87 & 13.87 &$ 2.151\times 10^{-13}$ & 0.18503\\
Her X-1& 0.85 & 13.548 &$ 2.255\times 10^{-13} $ & 0.18508\\
\hline
\end{tabular}
\end{table}
\subsection{Surface Redshift}
The gravitational redshift $ Z_{s} $ within a static line element can be obtained as 
\begin{eqnarray}
Z_{s}=\sqrt{g_{tt}(R)}-1=\sqrt{1-\frac{2M}{R}}-1\label{43}
\end{eqnarray}
where $ g_{tt}(R)=e^{\nu(R)}=1-\frac{2M}{R}$\\
\begin{figure}[h]
\begin{center}
\includegraphics[width=8cm]{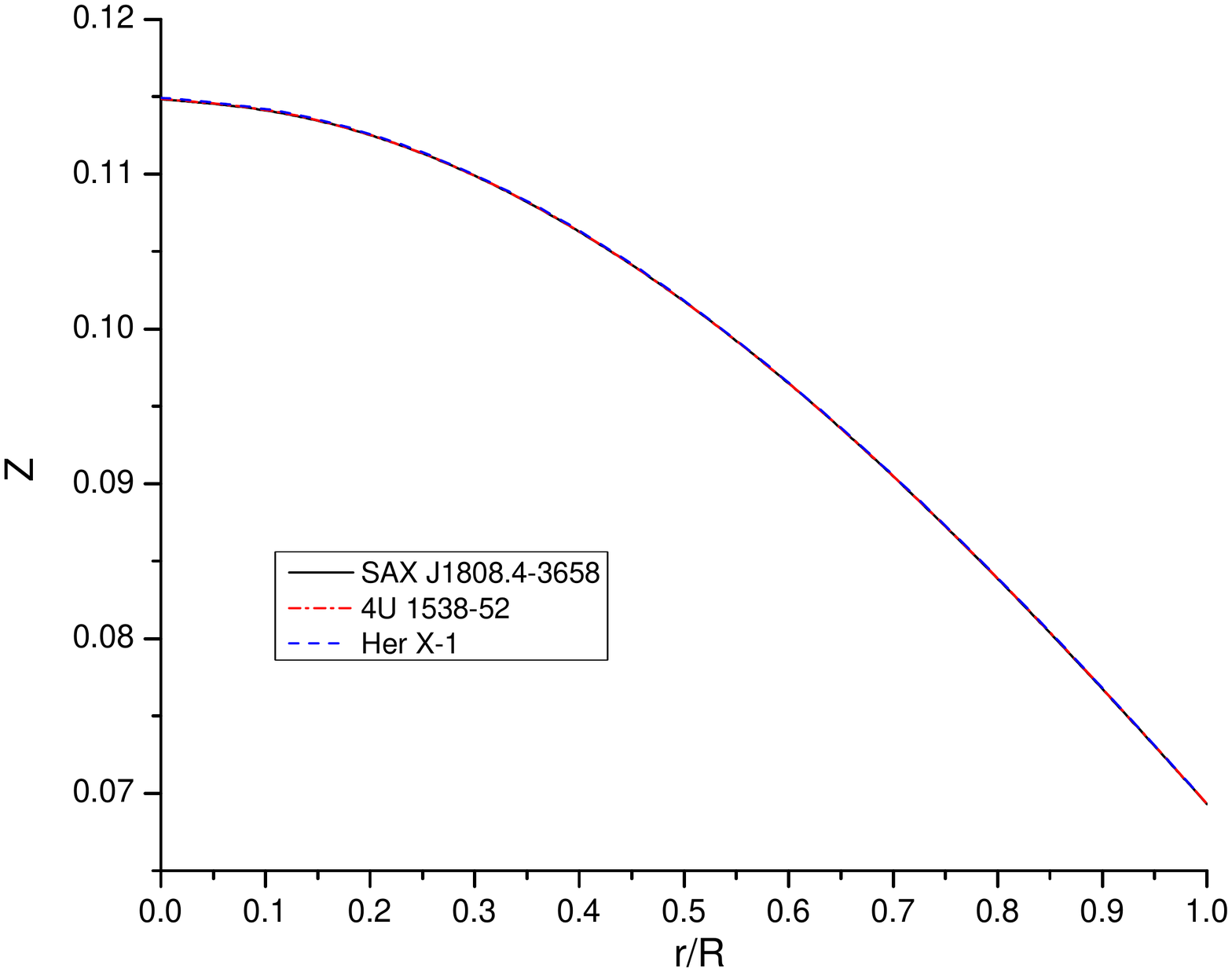}
\caption{Behaviour of Redshift vs. fractional radius
r/R for SAX J1808.4-3658, 4U 1538-52 and Her X-1. For plotting this figure we have employed data set values of physical parameters and constants which are the same as used in Fig.\ref{f2} }\label{f7}
\end{center}
\end{figure}
\begin{table}[h]
\caption{Energy densities, central pressure and Buchdahl condition for different compact star candidates for the above parameter values of Tables \ref{T1}.}
\label{T2}
\begin{tabular}{cccc}
\hline\rule[-1ex]{0pt}{3.5ex}
Compact star candidates& Central Density & Surface density& Central pressure\\
&$(gm/cm^{3})$&$(gm/cm^{3})$&$(dyne/cm^{2})$\\
\hline\rule[-1ex]{0pt}{3.5ex}
SAX J1808.4-3658& $1.383\times 10^{14}  $& $ 7.873\times 10^{13}  $ &$ 4.422\times 10^{34}$\\
4U 1538-52 & $1.48\times 10^{14}  $& $ 8.428\times 10^{13}  $ &$ 4.304\times 10^{34}$\\
Her X-1& $1.552\times 10^{14}  $& $ 8.836\times 10^{13}  $ &$ 0.452\times 10^{34} $\\
\hline
\end{tabular}
\end{table}
The maximum possible value of redshift should be at the center of the star and decrease with the increase of radius.Buchdahl\cite{Buchdahl} and Straumann\cite{straumann} have shown that for an isotropic star the surface redshift $ Z_{s}\leq 2 $.For an anisotropic star Bohmer and Harko \cite{bohmer} showed that the surface redshift could be increased up to $ Z_{s}\leq 5 $. Ivanov\cite{Ivanov} modified the maximum value of redshift and showed that it could be as high as $ Z_{s}= 5.211 $. In this model we have  $ Z_{s}\leq 1 $ for compact stars SAX J1808.4-3658,4U 1538-52 and Her X-1. Also it is decreasing towards the boundary(see Fig.\ref{f7}).
\begin{table}[h]
\caption{The numerical values of physical parameters of the star SAX J1808.4-3658 for $ C=2.009\times10^{-13}km^{-2}, K=7/4$}
\label{T3}
\begin{tabular}{cccccc}
\hline\rule[-1ex]{0pt}{3.5ex}
$ r/R $&$p(km^{-2})$&$\rho(km^{-2})$& $ p/\rho $ &$dp/d\rho$& Redshift\\
\hline\rule[-1ex]{0pt}{3.5ex}
0&	$ 2.66225\times10^{-5} $&	0.000823&	0.032363&	0.098283&	0.114842\\
0.1&	$ 2.62111\times10^{-5} $&	0.000817&	0.032083&	0.097498&	0.114301\\
0.2&	$ 2.50048\times10^{-5} $&	0.000801&	0.031238&	0.095115&	0.11269\\
0.3&	$ 2.30747\times10^{-5} $&	0.000774&	0.029807&	0.09106&	0.110046\\
0.4&	$ 2.05321\times10^{-5} $&	0.000739&	0.02776&	0.085206&	0.106427\\
0.5&	$ 1.75132\times10^{-5} $&	0.000699&	0.025056&	0.077382&	0.10191\\
0.6&	$ 1.41603\times10^{-5} $&	0.000654&	0.021649&	0.067378&	0.096586\\
0.7&	$ 1.06171\times10^{-5} $&	0.000607&	0.017489&	0.054947&	0.090557\\
0.8&	$ 7.00901\times10^{-6} $&	0.000559&	0.012525&	0.039815&	0.083927\\
0.9&	$ 3.44264\times10^{-6} $&	0.000513&	0.00671&	0.021685&	0.076806\\
1&	0&	0.000468&	0&	0.000246&	0.069299\\
\hline
\end{tabular}
\end{table}
\subsection{Equation of state (EOS)}
The equation of state(EoS),i.e. a relation between pressure and density is an important features of neutron star. So in this section we have discuss about EoS.It is wroth-wile to mentioned that different equation of state (EoS) of the neutron star lead to different mass-radius (M-R) relation. Many authors \cite{Dey1998,harko2002,Gondek2000} have suggested that the EoS $P=P(\rho)$ can be well approximated by a linear function of the energy density $\rho$ of compact star at high densities. Some researchers have also expressed more approximated form of the equation of state (EOS) $P=P(\rho)$ as a linear function of energy density $\rho$ (in details see  \cite{Haensel1989,Frieman1989,Prakash1990}). Here we find the EoS in a linear function of form $P=P(\rho)$ as, 
\begin{eqnarray}\label{EOS1}
\kappa\,p_r(r)&=& \frac{2\,(7+4\,{\tilde{\rho_1}} )}{7\,G(\rho_1)\,(1+ {\tilde{\rho_1}})} \times 
  \left[f_2(\rho_1)\, \Bigg(~ G_1(\rho_{1s})\,\bigg|\phi(\rho_1)\bigg|^{0.25}
 +\,\bigg|\phi(\rho_1)\bigg|^{-1.25}~\Bigg)
   -f_{1}(\rho_1)\,f_{3}(\rho_1)\,f_{4}(\rho_1) \,\right]\,\nonumber \\ &&-\frac{3C}{7\,(1+{\tilde{\rho_1}})},
\end{eqnarray}

where,\\\\
$ G(\rho_1)= \bigg[\frac{4 (1+\tilde{\rho_1})}{3}\bigg]^{\frac{1}{4}} [f(\rho_1)+1)]\,\Bigg[~ G_1(\rho_{2s})\,\bigg|\phi(\rho_1)\bigg|^{0.25} + \bigg|\phi(\rho_1)\bigg|^{-1.25}~\Bigg],\\\\
f_{2}(\rho_1)=\dfrac{2C}{3}
\Bigg[\big(f(\rho_1)+1\big)\Bigg(\frac{4(1+\tilde{\rho_1})}{3}\Bigg)^{-3/4}+2\Bigg(\frac{4(1+\tilde{\rho_1})}{3}\Bigg)^{1/4}f(\rho_1)\Bigg],\\\\
f_{3}(\rho_1)=\Bigg[G_{1}(\rho_{1s})(0.25)\bigg|\phi(\rho_1)\bigg|^{(-0.75)} +(-1.25)\bigg|\phi(\rho_1)\bigg|^{(-2.25)} \Bigg],~~~f_{4}(\rho_1)=\dfrac{-8C}{3f(\rho_1)\big(f(\rho_1)-1\big)^2} \\
f_{1}(\rho_1)=\Bigg(\frac{4(1+\tilde{\rho_1})}{3}\Bigg)^{1/4}\big(f(\rho_1)+1\big),~~~~~~~\phi(\rho_1)=\dfrac{5+2\tilde{\rho_1}+3f(\rho_1)}{2(1+\tilde{\rho_1})},\\\\
\tilde{\rho_1}=\frac{(1-2\,\rho_1) \pm \,\sqrt{1+8\,\rho_1}}{2\,\rho_1},~~~ \rho_1=\frac{7\,\kappa\,\rho}{3\,C},~~~f(\rho_1)=\sqrt{\frac{7+4\,\tilde{\rho_1}}{3}} ,\\\\
G_1(\rho_{1s})=\frac{f_2(\rho_{1s})\,L_2(\rho_{1s})\,\bigg|\phi(\rho_{1s})\bigg|^{-5/4}-\frac{L(\rho_{1s})}{f^2(\rho_{1s})}\,\bigg|\phi(\rho_{1s})\bigg|^{-5/4} + S(\rho_{1s})}{\frac{L(\rho_{1s})}{f^2(\rho_{1s})}\,\bigg|\phi(\rho_{1s})\bigg|^{1/4}-f_2(\rho_{1s})\,L_2(\rho_{1s})\,\bigg|\phi(\rho_{1s})\bigg|^{1/4} +P(\rho_{1s})},\\\\
 L(\rho_{1s})=\bigg[\frac{4 (1+\tilde{\rho_{1s}})}{3}\bigg]^{\frac{1}{4}} [f(\rho_{1s})+1],~~~L_2(\rho_{1s})=\frac{8\,f(\rho_{1s})}{9\,(1+\tilde{\rho_{1s}})} \\\\
f_{2}(\rho_{1s})=\dfrac{2C}{3}
\Bigg[\big(f(\rho_{1s})+1\big)\Bigg(\frac{4(1+\tilde{\rho_{1s}})}{3}\Bigg)^{-3/4}+\,2\Bigg(\frac{4(1+\tilde{\rho_{1s}})}{3}\Bigg)^{1/4}f(\rho_{1s})\Bigg], ~~~\\\\ P(\rho_{2s})=f_{1}(\rho_{2s})\,L_2(\rho_{1s})\,(-1/4)\,\bigg|\phi(\rho_{1s})\bigg|^{-3/4},~~~~\phi(\rho_{1s})=\dfrac{5+2\tilde{\rho_{1s}}+3f(\rho_{1s})}{2(1+\tilde{\rho_{1s}})}, \\\\
  f_{1}(\rho_{1s})=2\,\bigg[\frac{4 (1+\tilde{\rho_{1s}})}{3}\bigg]^{\frac{1}{4}}\,\frac{[f(\rho_{1s})+1]}{[f(\rho_{1s})-1]^2}, ~~~
 S(\rho_{1s})=f_{1}(\rho_{1s})\,L_2(\rho_{1s})\,(7/4)\,\bigg|\phi(\rho_{1s})\bigg|^{-9/2},\\\\
 \tilde{\rho_{1s}}=\frac{(1-2\,\rho_{1s}) \pm \,\sqrt{1+8\,\rho_{1s}}}{2\,\rho_{1s}},~~~  \rho_{1s}=\frac{7\,\kappa\,\rho_s}{3\,C},~~~f(\rho_{2s})=\sqrt{\frac{7+4\,\rho_{1s}}{3}}. $
 $\rho$ and $\rho_s$ \\ \\
\begin{figure}[h!]
\begin{center}
\includegraphics[width=6cm]{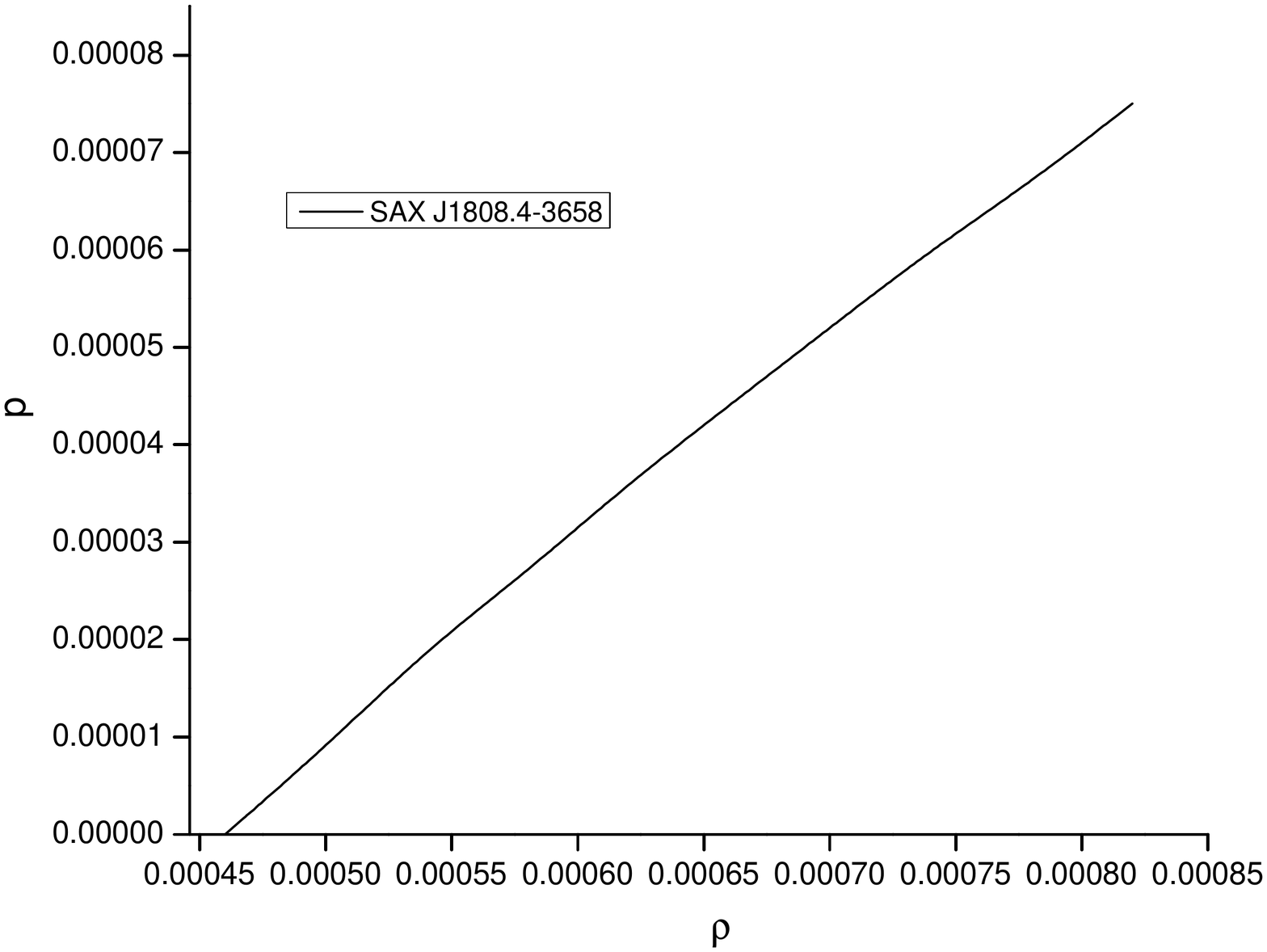}~\includegraphics[width=6cm]{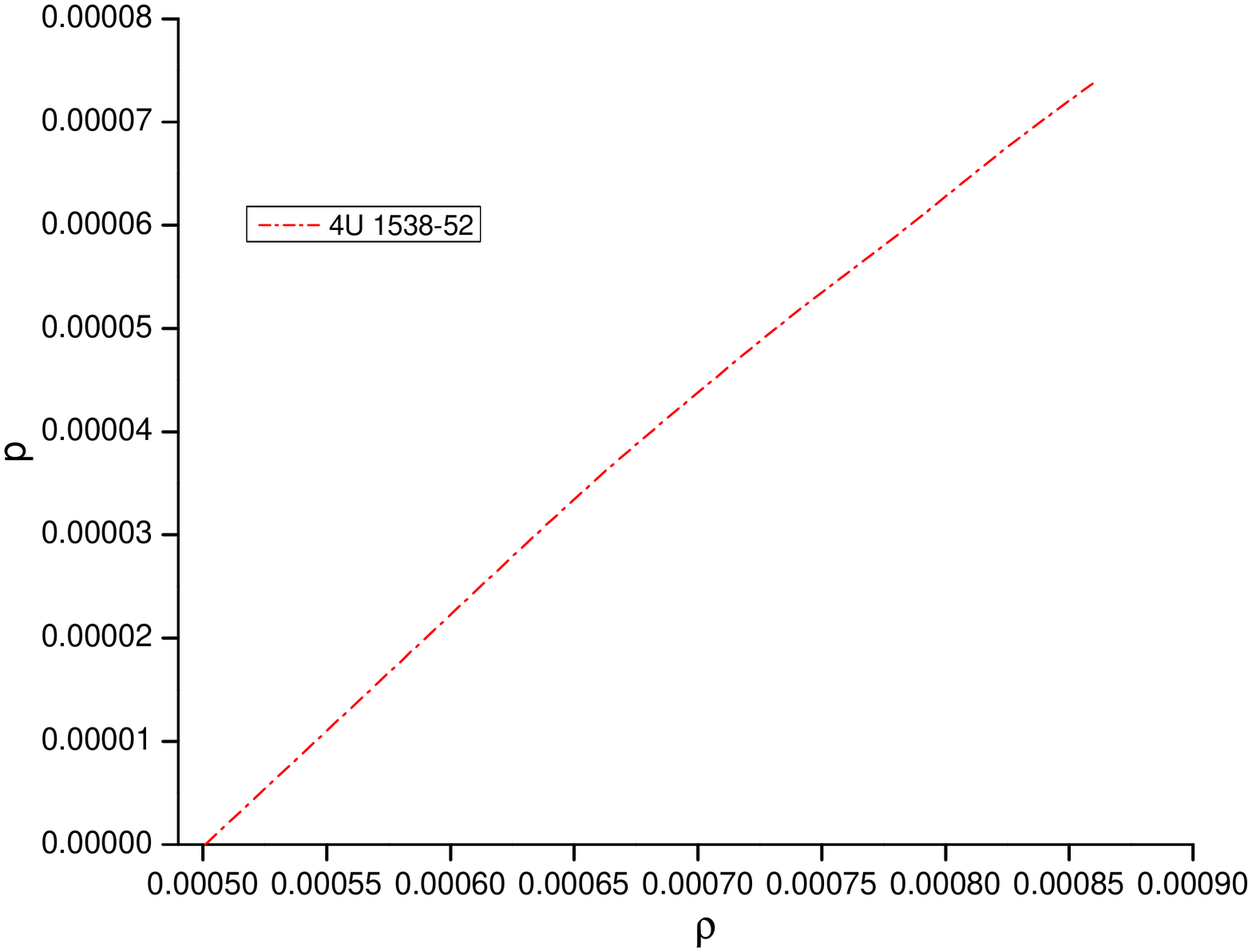}~\includegraphics[width=6cm]{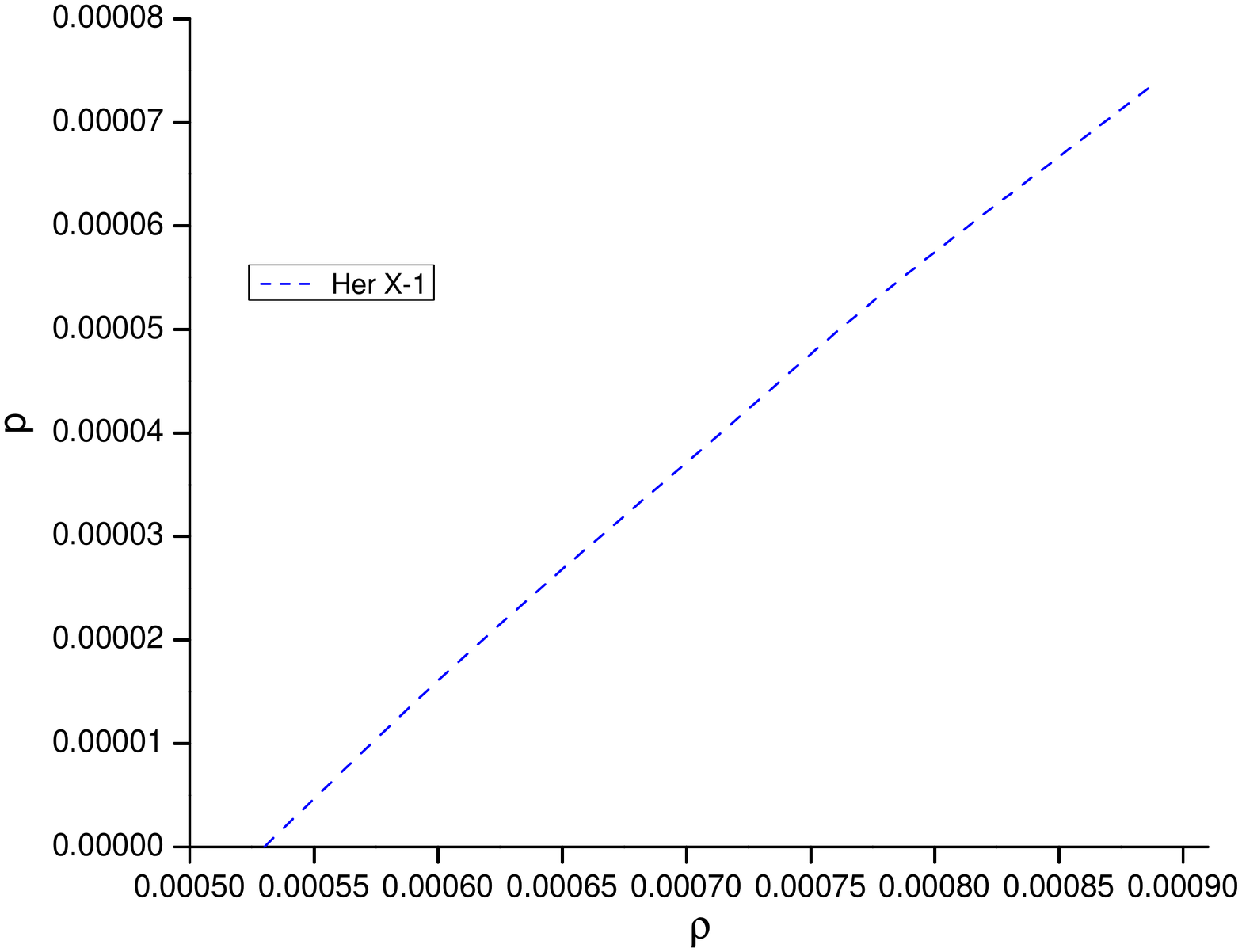}
\caption{Behaviour of pressure$p$ (in $km^{-2}$) vs. energy density$\rho$ ($km^{-2}$) for SAX J1808.4-3658, 4U 1538-52 and Her X-1 }\label{f8}
\end{center}
\end{figure}
From equation(\ref{EOS1}), we can observe that the pressures is a function of density, which describe the an EoS for SAX J1808.4-3658,4U 1538-52 and Her X-1. In an argument, Dey et al \cite{Dey1998} have proposed a new type of EoSs mainly describe strange matter. This was later generalised by Gondek-Rosinska et al\cite{Gondek2000} in a linear function of density ($\rho$), as
\begin{eqnarray}
p = \alpha\,(\rho-\rho_s),
\label{EOS}
\end{eqnarray}
where $\rho_s$ denotes surface density and $\alpha$ is non-negative constant. Harko and Cheng \cite{harko2002} have demonstrated that the equation(\ref{EOS}), gives the maximum mass of a strange star which is $M_{max}$ = $1.83 M_{\odot}$ when $\rho_s = 4B$ ( $B= 56 MeV fm^{3}$). In the present paper, we have developed same relation as considered by \cite{Gondek2000}. In that work, we have showed that the equation(\ref{EOS}) corresponds to self-bound matter at the surface density $\rho_s$. Fig.\ref{f8} represents the behavior of pressure verses density for compact stars with realistic EoS . In the Fig. \ref{f8}, we observe that the pressure $p$ vanishes at surface density $\rho_s$ i.e. at the boundary of our model. This implies that $p$ can be expressed by interpolation in power of $\rho-\rho_s$. Such parametrization is very convenient for stellar modelling, which also significant to the interior of stable stellar configurations \cite{Gondek2000}. 
\begin{table}
\caption{The numerical values of physical parameters of the star 4U 1538-52 for $ C=2.151\times10^{-13}km^{-2}, K=7/4$}
\label{T4}
\begin{tabular}{cccccc}
\hline\rule[-1ex]{0pt}{3.5ex}
$ r/R $&$p(km^{-2})$&$\rho(km^{-2})$& $ p/\rho $ &$dp/d\rho$& Redshift\\
\hline\rule[-1ex]{0pt}{3.5ex}
0&	$ 2.85086\times10^{-5} $&	0.000881&	0.032368&	0.098284&	0.114866\\
0.1&	$ 2.80682\times10^{-5} $&	0.000875&	0.032089&	0.097499&	0.114325\\
0.2&	$ 2.67753\times10^{-5} $&	0.000857&	0.031244&	0.095116&	0.112714\\
0.3&	$ 2.47093\times10^{-5} $&	0.000829&	0.029813&	0.09106&	0.110069\\
0.4&	$ 2.19861\times10^{-5} $&	0.000792&	0.027765&	0.085204&	0.106449\\
0.5&	$ 1.8753\times10^{-5} $&	    0.000748&	0.025061&	0.077378&	0.101931\\
0.6&	$ 1.5164\times10^{-5} $&	0.000701&	0.021653&	0.067371&	0.096606\\
0.7& $ 1.13697\times10^{-5} $&	0.000649&	0.017492&	0.054937&	0.090574\\
0.8&	$ 7.50583\times10^{-6} $&	0.000599&	0.012527&	0.039801&	0.083944\\
0.9&	$ 3.6867\times10^{-6} $&	0.000549&	0.006711&	0.021666&	0.076821\\
1&	0&	0.000502&	0&	0.000221&	0.069312\\
\hline
\end{tabular}
\end{table}
\subsection{Static stability criterion}
The most important feature of stability for stellar configuration is static stability criterion \cite{harrison,zeldovich}. In this criterion, it is postulate that the any stellar configuration has an increasing mass with increasing central density, i.e. $ dM/d\rho_{0} > 0 $ represents stable configuration and vice versa. If the mass remains constant with increasing central density, i.e. $ dM/d\rho_{0} = 0 $ we get the turning point between stable and unstable region. For this model, we obtained $ M(R) $ and $ dM/d\rho_{0} $ as follows-
\begin{figure}[h!]
\begin{center}
\includegraphics[width=7cm]{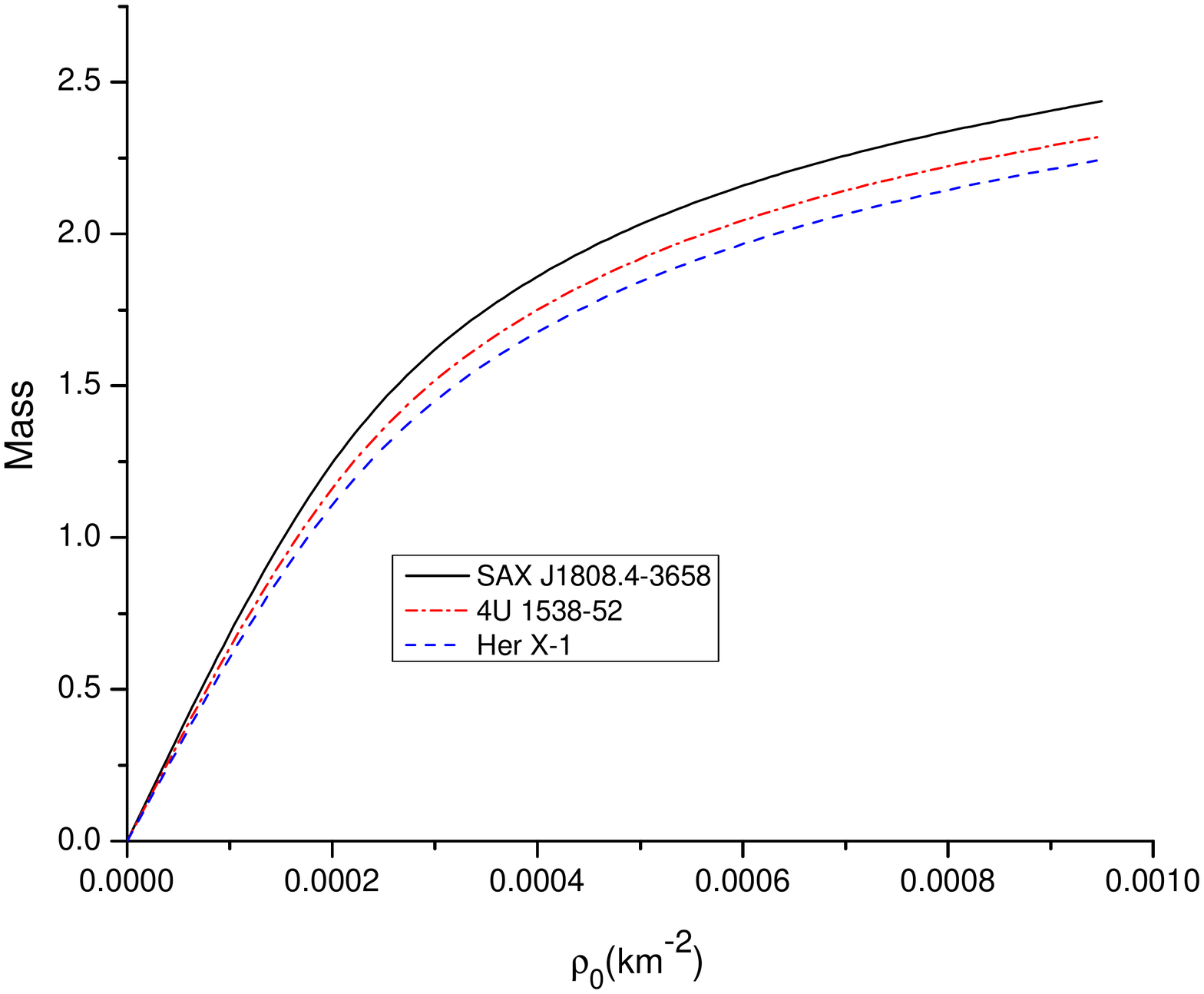}~\includegraphics[width=7cm]{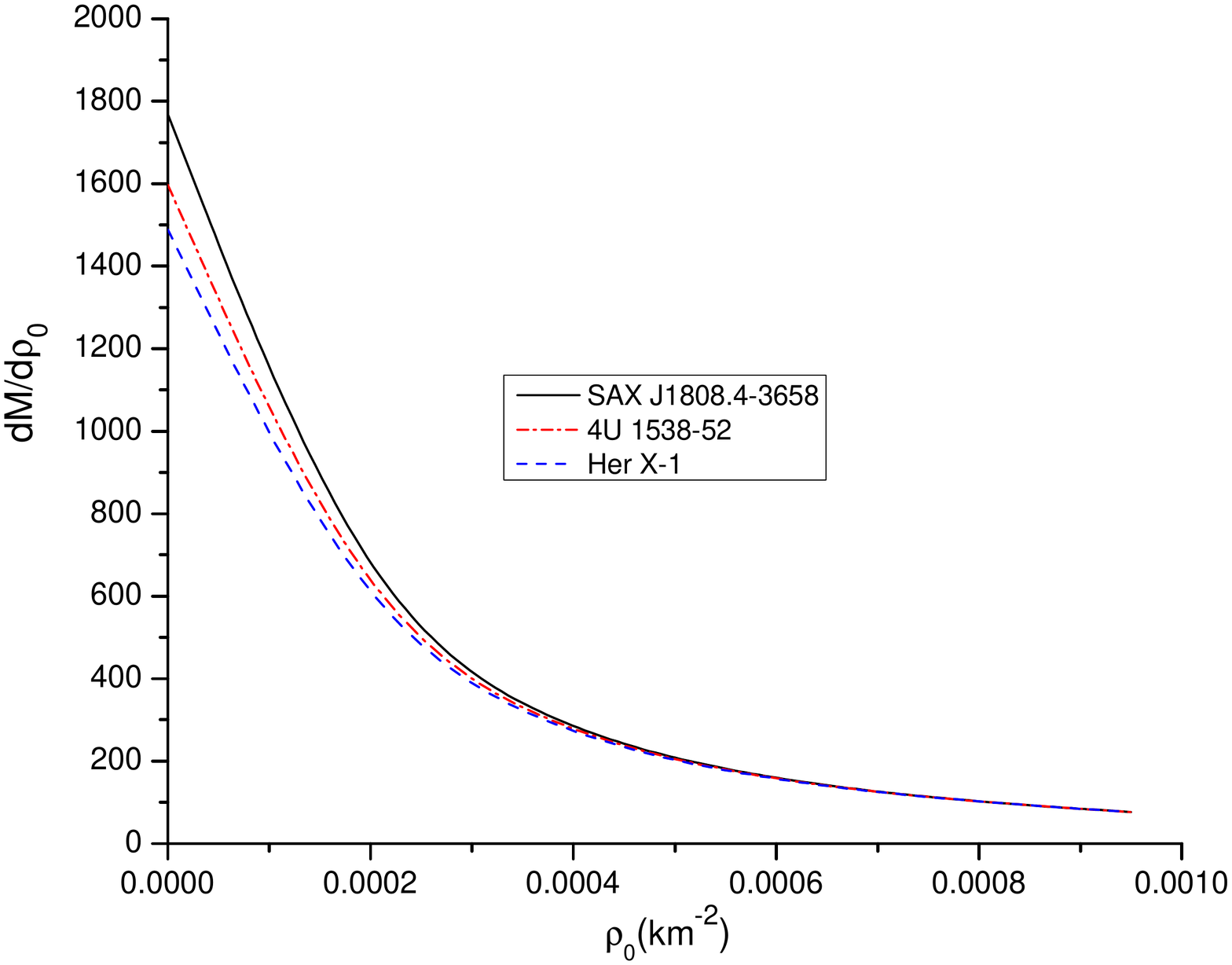}
\caption{Behaviour of Mass vs. central density(left) and $ dM/d\rho_{0} $ vs. central density(right) for SAX J1808.4-3658, 4U 1538-52 and Her X-1}\label{f9}
\end{center}
\end{figure}
\begin{equation}
M(R)=\frac{12\,\pi\,\rho_0\, R^3}{(9+56\,\pi\,\rho_0\,R^2)}~~~~~\textrm{and}~~~~\frac{dM}{d\rho_0}=\frac{108\,\pi\, R^3}{(9+56\,\pi\,\rho_0\,R^2)^2}\label{M11}
\end{equation}
Hence from Fig.\ref{f9},we can conclude that presenting model represents static stable configuration.
\begin{table}
\caption{The numerical values of physical parameters of the star 4U 1538-52 for $ C=2.151\times10^{-13}km^{-2}, K=7/4$}
\label{T5}
\begin{tabular}{cccccc}
\hline\rule[-1ex]{0pt}{3.5ex}
$ r/R $&$p(km^{-2})$&$\rho(km^{-2})$& $ p/\rho $ &$dp/d\rho$& Redshift\\
\hline\rule[-1ex]{0pt}{3.5ex}
0&	$ 2.99041\times10^{-5} $&	0.000924&	0.03238&	0.098287&	0.114915\\
0.1&	$ 2.94426\times10^{-5} $&	0.000917&	0.0321&	0.097501&	0.114373\\
0.2&	$ 2.80875\times10^{-5} $&	0.000899&	0.031255&	0.095117&	0.112761\\
0.3&	$ 2.59187\times10^{-5} $&	0.000869&	0.029824&	0.091059&	0.110115\\
0.4&	$ 2.30627\times10^{-5} $&	0.000831&	0.027776&	0.085201&	0.106493\\
0.5&	$ 1.96706\times10^{-5} $&	0.000785&	0.02507&	0.077371&	0.101972\\
0.6&	$ 1.59055\times10^{-5} $&	0.000734&	0.021661&	0.067359&	0.096644\\
0.7&	$ 1.19252\times10^{-5} $&	0.000682&	0.017498&	0.054918&	0.09061\\
0.8&	$ 7.87207\times10^{-6} $&	0.000628&	0.012532&	0.039773&	0.083976\\
0.9&	$ 3.86576\times10^{-6} $&	0.000576&	0.006714&	0.021628&	0.076851\\
1&	0&	0.000526&	0&	0.000171&	0.069338\\
\hline
\end{tabular}
\end{table}
\section{Conclusion}
In this article, we have discuss a new solution of Vaidya-Tikekar model for spherically symmetric uncharged fluid ball and found,it is physically valid solution.The Fluid ball contain an uncharged perfect fluid matter and Schwarzschild exterior metric. Mainly, we perform a detailed investigation of the physical result of high density system like uncharged fluid \cite{Lake,Klake,Herrera} and observe that the physical viability and acceptable of the our model in connection with compact star like Her X-1,4U 1538-52 and SAX J1808.4-3658.\par
It has been observe that the energy density and pressure are positive at the center i.e $ \rho_{0}>0, p_{0}>0 $ and monotonically decreasing throughout the fluid ball,see Fig.\ref{f2}. The energy conditions are very important to understand many theorem of general relativity such as singularity theorem of stellar collapse.
  Fig.\ref{f5} shows that the energy conditions are positive throughout the star and model satisfy (i) strong energy condition(SEC) and (ii) weak energy condition (WEC)\cite{YK}.We have also studied about the surface redshift. It should be  maximum at the center and monotonically decreasing from the center to surface\cite{Ivanov} see Fig.\ref{f7}. The modified TOV equation describes the equilibrium condition see Fig.\ref{f3} and observe that the gravitational force is balanced by the hydrostatic force. For stability analysis the adiabatic constant($ \Gamma $) is an important physical parameter and compact star will be stable if $ \Gamma>4/3 $\cite{Heint}. Fig.\ref{f6} show that $\Gamma > 4/3$, so model developed in this paper is stable.The mass-radius relation must be less than 8/9 \cite{Buchdahl}. Our model also satisfy this condition. The numerical values of physical quantities are shown in the Table \ref{T1}-\ref{T5}.We have obtained the EoS for the present compact star model, which is the significant physical property to describe structure of any realistic
  matter. We can see from equation(\ref{EOS1}) the pressure is purely function of density. Hence we conclude that this approach may help to describe the structure of compact star.

\end{document}